\documentstyle[12pt]{article}

\begin{document}


\hsize=6.15in
\vsize=8.2in
\hoffset=-0.42in
\voffset=-0.3435in

\normalbaselineskip=24pt\normalbaselines

{\it Journal of Theoretical Biology,} in press.

\vspace{1.5cm}

\begin{center}
{\large \bf  Conservation rules, their breakdown, and optimality in 
{\it Caenorhabditis} sinusoidal locomotion}
\end{center}

\vspace{1.2cm}

\begin{center}
Jan Karbowski$^{1,2,*}$, Christopher J. Cronin$^{1}$, 
Adeline Seah$^{1}$, Jane E. Mendel$^{1}$, \\
Daniel Cleary$^{1}$, and Paul W. Sternberg$^{1}$ 
\end{center}

\vspace{1.8cm}

\begin{center}
{\it $^{1}$ Howard Hughes Medical Institute and
Division of Biology 156-29, \\
$^{2}$ Sloan-Swartz Center for Theoretical Neurobiology,
Division of Biology 216-76, \\
California Institute of Technology,
Pasadena, CA 91125, USA }
\end{center}

\vspace{3.5cm}

\noindent $^{*}$ Corresponding author: jkarb@its.caltech.edu.\\
Phone: 626-395-5840; Fax: 626-568-8012;

\newpage

\begin{abstract}
Undulatory locomotion is common to nematodes as well as to limbless
vertebrates, but its control is not understood
in spite of the identification of hundred of genes involved in 
{\it Caenorhabditis elegans} locomotion. To reveal the mechanisms of 
nematode undulatory locomotion, we quantitatively analyzed the movement of 
{\it C. elegans} with genetic perturbations to neurons, muscles, and 
skeleton (cuticle). We also compared locomotion of different 
{\it Caenorhabditis} species. We constructed a theoretical model that 
combines mechanics and biophysics, and that is constrained by the 
observations of propulsion and muscular velocities, as well as wavelength 
and amplitude of undulations. We find that normalized wavelength is 
a conserved quantity among wild-type {\it C. elegans} individuals, 
across mutants, and across different species. The velocity of forward 
propulsion scales linearly with the velocity of the muscular wave and 
the corresponding slope is also a conserved quantity and almost optimal; 
the exceptions are in some mutants affecting cuticle structure. 
In theoretical terms, the optimality of the slope is equivalent to the 
exact balance between muscular and visco-elastic body reaction bending 
moments. We find that the amplitude and frequency of undulations 
are inversely correlated and provide a theoretical explanation
for this fact. These experimental results are valid both for young adults 
and for all larval stages of wild type {\it C. elegans}. In particular,
during development, the amplitude scales linearly with the wavelength, 
consistent with our theory. We also investigated the influence of
substrate firmness on motion parameters, and found that it does not affect
the above invariants. 
In general, our biomechanical model can explain the observed 
robustness of the mechanisms controlling nematode undulatory locomotion.
\end{abstract}

\vspace{1cm}


\newpage

\noindent {\bf Introduction}

Undulatory locomotion enables invertebrates such as nematode worms and
leeches, and limbless vertebrates such as snakes and fish to move through 
different environments including soil, sand, water, and tissues of plants 
in search of food (Gray, 1964).
Nematode body undulations are controlled by the neuromuscular system, 
which generates a wave of longitudinal muscle contractions modulated 
by elastic properties of the cuticle and hydrostatic skeleton.
Such a wave consists of alternating phases of dorsal and ventral muscle
contractions (worms lie on their sides) that travel posteriorly along 
the body length if the worm moves forward, and anteriorly if the worm moves 
backward. As a result of the interactions between neurons, muscles, 
skeleton/cuticle, and the environment, the worm crawls and its body follows 
approximately sinusoidal trajectory (Fig. 1).

There remains a major gulf between our understanding of biomechanics 
(Alexander and Goldspink, 1977; Gray and Lissmann, 1964; Niebur and Erdos,
1991; Wu, 1971; Cheng et al, 1998; Hirose, 1993; Ijspeert, 2001)  
and molecular genetics (Bargmann, 1998; Francis et al, 2003) of undulatory 
movement. Studies of leech and lamprey 
(Friesen and Cang, 2001; Skinner and Mulloney, 1998;
Lockery and Sejnowski, 1992; Marder and Calabrese, 1996; Williams, 1998; 
Cortez et al, 2004)
have led to a system-level, mostly neural, understanding of such 
movement but have little connection to how such movement is specified by 
the genome. Molecular genetic studies have identified hundreds of genes 
involved in locomotory behavior in {\it C. elegans} and while functional 
connections among many of these genes have been elucidated, they have not 
yet explained how the system works. Sensory behavior has begun to yield 
to a molecular approach (de Bono and Maricq, 2005), but remains a long way 
from motor output. We seek 
to understand how coordinated motor output is specified by a genome, and
specifically, how genes influence the parameters that directly control the
locomotion. We therefore started a two-pronged approach to this general 
problem in {\it C. elegans}. First, we started with single gene 
perturbations and measured behavior quantitatively. Second, we started 
building relatively simple models of worm movement, matching main parameters 
to experimental observables. This interactive approach allowed us to
construct a biomechanical model that fits experimental data on the 
characteristics of sinusoidal locomotion.

The experimental approach consisted of genetic perturbations to parameters 
relevant for locomotory control in {\it C. elegans\/} that included neurons, 
muscles, and cuticle, and quantitative data analysis of the resulting motion. 
We also compared locomotion of several wild-type {\it Caenorhabditis\/} 
nematode species to provide additional clues about generic characteristics. 
Our theoretical model combines mechanical and biophysical aspects of
undulatory locomotion. This model allows us to explain why some 
characteristics change or stay constant as a genetic perturbation is applied. 
In this respect, our model has a potential to provide richer information 
than standard, neural-level approaches in leech and lamprey 
(Friesen and Cang, 2001; Skinner and Mulloney, 1998;
Lockery and Sejnowski, 1992; Marder and Calabrese, 1996; Williams, 1998).

We find that although several quantities characterizing the movement 
vary from mutant to mutant, their inter-relationships are preserved in most 
cases, suggesting that locomotory control mechanisms are robust 
and somehow evolutionarily optimized. In particular, we find that our 
theory is consistent with the experimental findings that (i) the velocity 
of forward locomotion scales linearly with the velocity of the muscular 
contractions wave such that the former is close to optimal for 
all species and almost all mutants we examined, 
(ii) the wavelength of undulations, normalized with respect to body length, 
is highly conserved across different related species and different mutants, 
and (iii) the amplitude of the undulations weakly
decreases with the frequency of the wave. These results are preserved
during different developmental stages of {\it C. elegans}, as well as
on different substrate firmness.

The system responsible for locomotion of {\it Caenorhabditis\/} worms 
comprises four main elements: nerve ring (head) neurons, motor neurons, 
complexes of muscles with skeleton/cuticle, and different mechano-sensory 
feedback loops that influence activity of motor neurons (Fig. 2). 
The precise interactions between these elements in producing
oscillatory locomotory output remain unknown. In this paper, we simply
assume that this system is capable of producing an oscillatory wave
of muscle contractions that propagates along the worm's body.
The main results and conclusions below are independent of the particular 
oscillatory mechanism. Construction of the model is described in the 
Theoretical Model section.

\newpage

\noindent {\bf Experimental procedures}

\noindent {\it Mutant and species selection.} \\
We examined the locomotion of several 
{\it C. elegans\/} mutant classes with defects in neural, muscular, 
and cuticle functions. For mutations affecting neurons we studied: 
MT2426 {\it goa-1(n1134)}, PS1762 {\it goa-1(sy192)}, and PS4498 
{\it egl-30(tg26)}, which alter G-protein alpha subunit genes responsible 
for regulating synaptic transmission (Mendel et al, 1995; Segalat et al
1995; Moghal et al, 2003), 
{\it cat-2(e1112)}, which encodes a tyrosine hydroxylase,
an enzyme required for biosynthesis of dopamine (Lints and Emmons, 1999)
and {\it cat-4(e1141)}, which encodes GTP cyclohydrolase I,
an enzyme required in the process of synthesis of dopamine and serotonin
(Kapatos et al, 1999). {\it cat-2} and {\it cat-4} affect activities of 
dopaminergic neurons (Sulston et al, 1975). The specific neuronal mutants 
chosen by us differ from a vast majority of other neuronal mutants because of
their known hyperactive locomotion. We want to investigate this interesting 
feature more quantitatively. For mutations affecting 
muscles we studied several strains carrying mutations in the head region 
of body wall muscle myosin {\it unc-54\/}: RW130 {\it unc-54(st130)}, 
RW132 {\it unc-54(st132)}, RW134 {\it unc-54(st134)}, 
RW135 {\it unc-54(st135)}, RW5008 {\it unc-54(s95)}, and BC347
{\it unc-54(s74)}; these mutations are hypothesized to alter 
the contraction-relaxation cycle of the myosin-actin 
crossbridge formation by increasing its duration (Moerman and Fire, 1997). 
For mutations affecting cuticle, we studied two loss-of-function alleles
of {\it sqt-1}: BE101 {\it sqt-1(sc101)\/} and BE103 {\it sqt-1(sc103)\/} 
mutants. {\it sqt-1\/} encodes a cuticle collagen (Kramer et al, 1988), 
a protein responsible for elastic and structural properties of the cuticle. 
We also studied a mutant BE109 of an unknown gene that also affects cuticle 
by removing the struts that separate layers of the cuticle 
(J. Kramer, pers. comm.), and two mutants with
increased body length: CB185 {\it lon-1(e185)\/} and double mutant 
PS697 {\it lon-1(e185); lon-2(e678)}. The {\it lon-1} gene encodes a 
protein belonging to the PR-protein superfamily that regulates 
polyploidization and body length (Brenner, 1974; Maduzia et al 2002). 
Moreover, we examined several 
wild-type nematode species closely related to N2 {\it C. elegans\/}
(Brenner, 1974): 
SB339 {\it C. japonica\/} (Kiontke et al, 2002), AF16 {\it C. briggsae\/}, 
SB146 {\it C. remanei\/} (A. Fodor, pers. comm.), CB5161 
{\it Caenorhabditis\/} sp. (A. Fodor, pers. comm.), and PS1010
{\it Caenorhabditis\/} sp (R. Giblin-Davis, pers. comm.).

\vspace{1cm}

\noindent {\it Description of experimental setup.} \\
We video recorded and digitized the motion of young adult hermaphrodite 
(or female from male-female strains) {\it Caenorhabditis} worms: wild-type 
{\it C. elegans} and their mutants, related species, and additionally 
{\it C. elegans} larvae. Young adults were 15-20 hr post mid-L4 
developmental stage, and larvae were in all stages from L1 to young adults. 
Adult {\it C. elegans} and related species are tiny animals with length 
about 1 mm and width about 0.1 mm, so our experimental set-up involved a
microscope connected to a video and a specialized open source software. 
Initially, an agar plate was 
covered with thin film of {\it E. coli} OP50 bacteria mixed with LB media. 
After that, the agar plate was covered with water (about 0.1 mm of hight) 
and left for 1 hour so that the mixture of bacteria with LB media 
is absorbed in the agar and the surface dry. Thus, during the recording 
sessions worms moved not through water but through a layer of bacteria with
some remaining LB media. That type of movement can be classified as
crawling, since the worms touch the substrate (agar), and are only partially 
immersed in the bacteria layer.

In experiments with different substrate firmness, the concentration of
agar in the substrate solution was changed. Higher agar concentration 
corresponds to a more stiff substrate on which worms move.

The video recording and data extraction was done using a device 
specially designed for studying {\it Caenorhabditis\/} locomotion 
(Cronin et al, 2005). We collected 5 minutes of video per worm, extracting
digital locomotion data from the middle 4 minutes.
Such 4 minute windows average over possible sensory influences that can 
vary among worms and thus statistically minimize the variability 
of external conditions. From these data we derived values of the velocity 
of forward and backward motion, frequency, wavelength, and amplitude of 
undulations. Our experimental setup allows us to measure both instantaneous 
and average values of crawling parameters.

\vspace{1cm}

\noindent {\bf Theoretical model}

The theoretical part of this study involved the construction of 
a comprehensive mathematical model, from which we could compute the 
quantities directly measured in the experimental part.

\noindent {\it Mechanics.}\\
The mechanical aspect of the undulatory locomotion is modeled by assuming
that worm's body can be treated as an active bending beam 
(Wu, 1971; Cheng et al, 1998). Crucial for the motion generation is the
existence of neuromuscular wave that propagates along worm's body. In
{\it C. elegans} such a wave can be generated by long-range interactions 
between body segments, as it was shown theoretically for a system of coupled 
oscillators (Kopell and Ermentrout, 1988). From a biological perspective,
long-range interactions could be mediated by stretch receptors, which 
are hypothesized (in White et al, 1986; Chalfie and White, 1988) to be 
located on endings of the extended dendritic processes of the excitatory 
motor neurons (Fig. 2). These extensions can be as long as 25 $\%$ of the
worm's length. For the forward motion B neurons, these processes extend
posteriorly, i.e., the coupling via stretch receptors comes from the 
posterior parts of the body. For the backward motion A neurons, the
extended processes are directed in the opposite direction, and consequently 
the stretch receptor coupling comes from the anterior parts of the body.
This distinction in dendrite directionality is correlated with observed
opposite directions of neuromuscular waves in both locomotory circuits.

Rhythmic muscle contractions, caused by the motor neurons'
activity, bend the worm's body and this generates the propulsion of its 
center of mass. Newtonian equations of motion describe the balance between
muscular, elastic, frictional, and internal pressure forces. 
During crawling, inertial effects are negligible because of the 
small mass of the worms (Gray, 1964) and small maximal acceleration
(for estimation see, Niebur and Erdos, 1991). Equations of motion are written
for a slice of a worm, perpendicular to its main body axis, and for a small 
slope of undulation. We choose a system of coordinates such that at 
resting conditions when the worm's body is straight, its main body axis is 
parallel to the x coordinate (Fig. 3). The transverse force balance is

\begin{equation}
- F + (F + \delta F)\cos(\delta\phi) + (G + \delta G)\sin(-\delta\phi)
- F_{N}\delta x = 0,
\end{equation}\\
and the longitudinal force balance is

\begin{equation}
- G + (G + \delta G)\cos(\delta\phi) - (F + \delta F)\sin(-\delta\phi)
+ F_{L}\delta x = 0,
\end{equation}\\
where $F$ is the beam shear force, $G$ is the longitudinal tension (it
includes both tension in structural elements of cuticle and internal
hydrostatic pressure), $\delta\phi$ is an infinitisemal change of 
undulations angle over the slice width, and $F_{N}$ and $F_{L}$ are 
the normal and longitudinal components of the frictional force per unit 
length of the slice. We assume that these components are proportional 
to the normal $v_{N}$ and tangential $v_{L}$ components of the animal's 
velocity, with proportionality coefficients $c_{N}$ and $c_{L}$, 
respectively. With the sign convention as in Fig. 3, we have 
$F_{N}= - c_{N}v_{N}$ and $F_{L}= c_{L}v_{L}$. This assumption is analogous 
to an assumption made in fluid dynamics models, and it is known as 
``Resistive Force Theory'' (Gray and Hancock, 1955; Lighthill, 1976). 
The normal and longitudinal components of velocity can be represented by
main body axis velocity $v$ (it is parallel to the main body axis), 
lateral velocity $u= {\partial h}/{\partial t}$ with $h$ being the lateral
displacement, and the tangential angle 
$\phi$ to the slice in the form: $v_{L}= v\cos\phi - u\sin\phi$, 
$v_{N}= v\sin\phi + u\cos\phi$.

After expanding $\cos(\delta\phi)$ and $\sin(\delta\phi)$ for small 
$\delta\phi$ in Eqs. (1) and (2), and neglecting higher order terms, 
we obtain the following differential equations of motion:

\begin{equation}
\frac{\partial F}{\partial x} - G\frac{\partial \phi}{\partial x}  
= - c_{N}v_{N}(x,t),
\end{equation}\\
and 

\begin{equation}
F\frac{\partial \phi}{\partial x} + \frac{\partial G}{\partial x} 
= - c_{L}v_{L}(x,t).
\end{equation}\\
The beam shear force $F$ is related to the total bending moment $M$ by

\begin{equation}
F= - {\partial M}/{\partial x}. 
\end{equation}

Equations (3)-(5) constitute the basis for our analysis of mechanics
of worm's undulatory locomotion. They can be further simplified in
the limit of small angle of undulations $\phi$. In this limit 
$\phi\approx \sin\phi \approx \tan\phi = {\partial h}/{\partial x}$ 
and $\cos\phi\approx 1 - \phi^{2}/2$, and as a result velocities 
take the forms:
$v_{L}\approx v\left[1-\frac{1}{2}({\partial h}/{\partial x})^{2}\right] 
- ({\partial h}/{\partial t})({\partial h}/{\partial x})$,
and 
$v_{N}\approx v{\partial h}/{\partial x} + {\partial h}/{\partial t}$. 
Also, the body curvature 
${\partial \phi}/{\partial x}\approx {\partial^{2}h}/{\partial x^{2}}
\approx 0$, and 
therefore, the term containing longitudinal tension $G$ in Eq. (3) can 
be neglected. Using this information, we can combine Eqs. (3) and (5) 
to arrive at the equation describing spatio-temporal pattern of the 
total bending moment:

\begin{equation}
\frac{\partial^{2} M(x,t)}{\partial x^{2}} \approx 
c_{N}\left[ v\frac{\partial h}{\partial x} 
+ \frac{\partial h}{\partial t} \right]. 
\end{equation}\\
One usually solves this equation for $M$ given some sinusoidal form
of the lateral displacement $h(x,t)$ that mimics undulatory locomotion
(Cheng et al, 1998). This is inverse to a seemingly more ``natural'' 
approach, in which some form of the bending moment would be chosen and 
the equation solved for $h(x,t)$. The general problem with the latter
approach is that it is hard to guess the right form of $M(x,t)$ which
produces stable undulatory motion. In particular, the spatial dependence of
$M$ is crucial for stabilization, and, as it turns out, $M$ depends
nontrivially on the position $x$ along the body (see below).
Also, it is not clear what
boundary conditions to impose on the $h(x,t)$ function, since during
real undulatory motion the animals' head and tail are in permanent motion.
For these reasons, we adopt the inverse approach of Cheng et al.
(Cheng et al, 1998) and solve Eq. (6) for $M$ with boundary conditions 
$M(0,t)= M(L,t)= \partial M(0,t)/\partial x =
\partial M(L,t)/\partial x = 0$, where $L$ is the worm's body length.
These conditions follow from the natural requirement that all forces and 
moments must disappear outside the borders of worm's body.
We choose the lateral displacement $h$ in the form:

\begin{equation}
h(x,t)= A_{0}\cos(\omega t - 2\pi x/\lambda) + \delta h(x,t),
\end{equation}\\
where $A_{0}$ is the amplitude of undulations, $\omega$ and $\lambda$ 
are the angular frequency and wavelength characterizing the neuromuscular 
wave which travels to the right with the velocity $\lambda\omega/2\pi$, 
and $\delta h(x,t)= A(t) + B(t)x$ is the so-called recoil correction 
(Cheng et al, 1998). This correction ensures that the total external 
force acting on worm's body is zero at all times
(Lighthill, 1960; Pedley and Hill, 1999).
From a mathematical point of view, this correction is necessary to
satisfy the boundary conditions, and the functions $A(t)$ and $B(t)$ are 
determined self-consistently from them. From a biological perspective, 
one can view this correction as a mechanosensory feedback contribution, 
which is necessary for coordination of the movement.

The total bending moment $M$ is composed of two additive contributions:
$M= M_{m} + M_{e}$, where the muscle generated bending moment $M_{m}$ is
given by Eq. (12) (see below), and visco-elastic reaction of the 
body moment $M_{e}$ is given by:

\begin{equation} 
M_{e} = - EJ\partial^{2}h/\partial x^{2} - \mu J\frac{\partial}{\partial t}
(\partial^{2}h/\partial x^{2}), 
\end{equation}\\
where $E$ is the Young's (elastic) modulus 
of the hydrostatic skeleton and cuticle, $J$ is the inertial moment of the 
body in relation to the axis perpendicular to the body midline, $\mu$ is the 
viscous coefficient characterizing relaxation processes in the hydrostatic
skeleton and cuticle.

The propulsion velocity is equivalent to the main body axis velocity 
$v$. Having the total bending moment $M$, we can find the 
equation for the propulsion velocity from Eqs. (4) and (5). First, we 
substitute Eq. (5) for the beam shear force $F$ in eq. (4), and then 
we use the fact that in the limit of a small slope of undulations 
$v_{L}\approx v\left[1 -\frac{1}{2}({\partial h}/{\partial x})^{2}\right] 
- ({\partial h}/{\partial t})({\partial h}/{\partial x})$.
Both corrections to $v_{L}$ or $v$ are small of the order of $O(h^{2})$,
however only the second, 
$({\partial h}/{\partial t})({\partial h}/{\partial x})$, is relevant
for the determination of $v$ in the $O(h^{2})$ order; the first correction
provides a higher order contribution $O(h^{4})$. This allows us to derive
$v$ from a resulting equation. However, because both $M$ and $h$ are 
functions of space and time, we must perform spatial and temporal
averaging of this equation. Spatial averaging is done within the limits
of 0 and $L$, where $L$ is the worm's length, while temporal averaging
is performed over one period of oscillations. After all the steps we
obtain:

\begin{equation}
v  \approx \frac{1}{L} \int_{0}^{L} dx \left[ \frac{1}{c_{L}}
\langle\frac{\partial M}{\partial x} \frac{\partial^{2}h}{\partial x^{2}}
\rangle +
\langle\frac{\partial h}{\partial t} \frac{\partial h}{\partial x}
\rangle \right],
\end{equation}\\
where the bracket $\langle...\rangle$ denotes temporal averaging, and
we neglected the contribution coming from the longitudinal tension, 
$\langle G(L)-G(0)\rangle$, because of a partial or complete cancelation 
of $\langle G(L)\rangle$ and $\langle G(0)\rangle$ in the limit of small 
slope of undulations.
Equation (9) allows us to determine self-consistently $v$, since
the right hand side of it also depends on $v$ (via $M$).
Note that both terms under the integral in Eq. (9) are of the same order,
i.e., $O(h^{2})$, since the total bending moment $M$ is of the order $O(h)$
(see Eq. (6)). This implies that the propulsion velocity $v$ is $O(h^{2})$
order quantity.

\vspace{0.5cm}

\noindent {\it Muscle dynamics.} \\
Muscle activity is modeled as a simple first-order kinetic equation,
in which an external driving factor activates muscles. That factor
originates from neural oscillations at the neuromuscular junction
and subsequent calcium influx. Since the magnitude of neural oscillations
can in general be position dependent (as is the total bending moment
$M$; see below) and we do not know in advance what that dependence could
be, in what follows, we will compute only space averaged (over the whole
worm's length) quantities associated with muscle activation.
We use the following equation:

\begin{equation}
\frac{d \overline{n_{\alpha}}(t)}{d t}= 
f \left[\overline{D_{\alpha}}(t)-\overline{n_{\alpha}}(t)\right] 
- g \overline{n_{\alpha}}(t), 
\end{equation}\\
where $\overline{n_{\alpha}}(t)$ is the space-averaged muscle activity level 
at time $t$, $\alpha$ denotes either dorsal (d) or ventral (v) side 
of the worm, $\overline{D_{\alpha}}(t)$ is the average driving factor, and
the parameters $f$ and $g$ correspond to the rates of muscle 
activation and deactivation, respectively. 
We choose the space-averaged driving factor $\overline{D_{\alpha}}(t)$ 
in the form:

\begin{equation}
\overline{D_{\alpha}}(t)= 
(T/2L)\left[1 \pm \cos(\omega t + \xi)\right],
\end{equation}\\
where the sign $+ (-)$ corresponds to dorsal (ventral) side, $L$ is the worm's
length, $T$ is the constant characterizing the overall
amount of calcium/synaptic transmission at the neuromuscular junction coming 
from neural oscillations, 
and $\xi$ is some phase factor reflecting temporal delays in neural and body 
reaction activities.

There are two equivalent ways of thinking about what constitutes muscle 
activity. From a biophysical perspective one can interpret muscle activity 
as the level of muscle contraction, which is related to the fraction
of crossbridges created (Huxley, 1957). In this model, myosin and actin 
molecules interact to create crossbridges between thick and thin filaments. 
The crossbridge dynamics result in both filaments sliding past one another, 
which leads to muscle contraction and force generation. In this 
interpretation, $f$ is the effective rate of crossbridge association and 
$g$ is its dissociation rate. Alternatively, from a physiological 
perspective, one can think about $\overline{n_{\alpha}}(t)$ as the 
position averaged muscle membrane potential.

Muscle dynamics and mechanics can be coupled by relating the muscle 
activity $\overline{n_{\alpha}}$ to the spatial average of the muscle 
bending moment $\overline{M_{m}}$. We assume that the average muscle 
bending moment $\overline{M_{m}}(t)$ is proportional to the difference 
in average activities of the dorsal $\overline{n_{d}}$ and ventral 
$\overline{n_{v}}$ muscles with a proportionality constant $\kappa$ 
measuring muscle stiffness, i.e., 

\begin{equation}
\overline{M_{m}}(t)= \kappa[\overline{n_{d}}(t)-\overline{n_{v}}(t)]. 
\end{equation}\\
This coupling allows us to derive the amplitude of undulations $A_{0}$ as 
a function of the frequency. This can be done by balancing spatially
averaged bending moments derived from the mechanics and muscle dynamics 
parts. The details are presented in the Appendix.

\vspace{2cm}

\noindent {\bf Experimental results}

\noindent {\it Primary locomotory data.} \\
In Tables 1-3 we present primary data we collected, which includes 
average values of the velocity of propulsion $v$, frequency of undulations
$\omega/2\pi$, amplitude of undulations $A_{0}$, wavelength of the body 
posture $\lambda$, the ratio of the wavelength to the length of the body
$\lambda/L$, and the ratio of the propulsion 
velocity to the velocity of the neuromuscular wave denoted by $\gamma$
(the efficiency coefficient; see below). Although, many of these parameters
change under different conditions, there are some regularities in the data.

Table 1 contains locomotory data for wild-type {\it C. elegans} during its
development through different larval stages to young adult. We observe that
worms' velocity increases monotonically as they mature and this is 
accompanied by monotonic changes in amplitude and wavelength, with 
frequency exhibiting some variability. Contrary to that trend, 
the body-length normalized wavelength $\lambda/L$ and the efficiency 
coefficient $\gamma$ stay relatively constant; about 2/3 for $\lambda/L$, 
and $\gamma$ in the range $0.8-0.9$.

In Table 2, we present data for adult {\it C. elegans} mutants and related
species, with wild-type {\it C. elegans} as a control. Most hyperactive 
mutants with affected nervous system move with greater velocity 
than wild-type. The outlier from this trend {\it cat-4} is an interesting 
case because it moves slower but with a higher frequency as compared 
to wild-type. It is important to note that our neuronal mutants are 
exceptions, because the majority of other neuronal mutants 
(not examined by us) are sluggish, which is the opposite behavior. 
Mutants affecting muscle structure and properties of the 
cuticle move slower than wild-type and hyperactive neuronal mutants. 
The locomotory data show high variability in
velocity, frequency, and amplitude across different mutants and species.
However, again, the normalized wavelength $\lambda/L$ and the efficiency 
coefficient $\gamma$ do not change much, especially $\lambda/L$ which
is around 2/3. In most cases the 
coefficient $\gamma$ is in the range $0.7-1.0$. There are
few exceptions with $\gamma\approx 0.5-0.6$. These include {\it cat-4},
{\it sqt-1}, and one strain of {\it unc-54}. The latter is probably a
consequence of the fact that these worms move very slow, often pausing,
with mean velocity comparable to its standard deviation.

Table 3 presents locomotory data for adult {\it C. elegans} on substrates
with different firmness (solidification) properties that can be changed by
agar content in the substrate solution. We observe that as substrate
becomes more firm, velocity and frequency decrease monotonically,
suggesting that it is harder for worms to move forward. Despite these
changes, the normalized wavelength is quite invariant, similar to the
results in Tables 1 and 2. The efficiency coefficient $\gamma$ also exhibits 
little variability. It is relatively constant with values in the range
$0.7-0.8$ up to $6 \%$ of agar concentration, and only for $8 \%$ it slightly 
drops. The value of $2 \%$ of agar concentration corresponds to a standard
value for which substrate is prepared and data in Tables 1 and 2 were
obtained for this particular value.

\vspace{0.5cm}

\noindent {\it Relationships between locomotory parameters.} \\
In order for the worms to move forward (or backward), the wave of 
neuro-muscular activity must travel backward (or forward). One of the main
characteristics of this wave is its wavelength. In Fig. 4 we plot the
normalized wavelength as a function of frequency, for a population of 
wild-type {\it C. elegans}, for different mutants, and for related 
{\it Caenorhabditis} species. As could be expected from the data in 
Tables 1-3, these results for the forward locomotion show that the normalized
wavelength $\lambda/L$ is practically frequency independent, and moreover,
its value is conserved across all three cases and is around 2/3.
For backward movement, we observe that $\lambda/L$ is also about $2/3$ 
(data not shown). This result is consistent with the similarities in the 
neuroanatomical structure of the forward and backward motion neural circuits
(White et al, 1986).

In Fig. 5, we investigate the relationship between the velocity of worm's 
propulsion and the velocity of the neuro-muscular wave given by 
$\lambda\omega/2\pi$. We plot the data 
for a population of wild-type {\it C. elegans}, for different mutants, and 
for related species. 
We find that these two velocities scale linearly with the proportionality 
coefficient (least-square fit slope) around 0.8 across all three conditions.
The fact that this coefficient is always $< 1$ implies
that the velocity of worm's propagation is less than the velocity of the
neuromuscular wave, and this is a consequence of the fact that 
{\it C. elegans} worms slip during sinusoidal motion. However, as can
be seen from Tables 1-3 and Fig. 5, the slippage, defined as $1-\gamma$,
is small and to a large extent independent of wavelength, frequency, 
velocity, amplitude, and even the firmness of the substrate. It also does 
not change during development.

Developmental data provide a broader parameter space for worm's body length
and other related parameters. In particular, we find that as worm's 
length increases, the amplitude and wavelength increase in a correlated 
manner. They are related by a linear scaling (Fig. 6), which implies that 
during development their ratio is conserved, i.e., $A_{0}/\lambda = 1/5$. 
However, for adult mutants and species, there is some variability in this
relationship, because the amplitude depends also on frequency and other 
biophysical parameters (see below).

A particular class of {\it unc-54} mutants studied by us have significantly
reduced frequency of undulations as compared to wild-type worm, and yet
they move with a slightly greater amplitude of the wave 
(Table 2; and for an extreme example, see Fig. 1, panel d). 
These data suggest that there may be some inverse relationship between the 
amplitude and frequency of undulations. To test this empirical observation, 
we determined the statistical significance of this inverse relationship 
by computing the correlation between 
the amplitude and frequency for the forward motion for the wild-type
{\it C. elegans} larvae and adults and on different substrate, for mutants,
and related species (Tables 4 and 5). We computed correlation coefficients 
for both instantaneous values of amplitude and frequency and their 
average values. Indeed, these two parameters are negatively correlated in 
all cases and for almost all worms with the same genotype. There are a few 
exceptions, but because the p-values in those cases are large,
they should be excluded as not statistically significant. In general,
correlations for instantaneous values of amplitude and frequency are smaller
but more statistically significant than those for their average values.
The lower negative correlations and lower p-values for instantaneous values 
probably reflect much larger sample size and consequently smaller variability 
in instantaneous locomotory parameters than in averaged parameters. 
Negative correlations are also present for worms with different genotypes: 
among different mutants of {\it C. elegans} including wild-type
the correlation coefficient is equal to $-0.293$, and among different 
{\it Caenorhabditis} species the correlation is of comparable value $-0.329$.
Similar negative correlations are present also for combined forward
and backward motion (data not shown).

\vspace{2cm}

\noindent {\bf Theoretical results and their relation to experimental
data}

What could be a possible explanation of the constancy of the normalized 
wavelength? It was suggested many years ago by L. Byerly and R. Russell 
(cited in White et al, 1986) that the wavelength may be related to the spatial 
extent of motor neuron's dendrites that have presumably stretch receptors 
at their endings (Fig. 2). These researchers concluded that since the 
dendrite length is fixed then so should be the wavelength. 
It can be shown, using a coupled 
oscillator model (Kopell and Ermentrout, 1988; Cohen et al. 1992; 
J. Karbowski, unpublished results), that the Byerly-Russell hypothesis 
provides only a partial explanation of this conservation phenomenon. 
In fact, the wavelength of the emerging wave depends not only on the spatial 
extent of long-range coupling but also on the nature 
of the interaction function, which in turn, depends on the underlying 
biophysics. The latter factor could be potentially frequency dependent
(Gutkin et al, 2005), but only for frequencies that are much larger, of
the order of 50 Hz, than relevant frequencies for nematode locomotion,
which are $\sim 1$ Hz. This probably can explain the independence of the 
wavelength on frequency for
wild-type {\it C. elegans} and among related species, which have similar
nervous systems. The matter is more subtle with {\it C. elegans}
mutants. It is natural to expect that mutations not directly affecting
neural function should not influence long-range coupling. On the
other hand, mutations investigated by us which do affect neural function,
affect only local inter-neuronal synaptic transmission. These mutations
do not affect dendritic structure, and thus, presumably do not alter
activities of stretch receptors. It thus follows that all our mutations
should not influence long-range coupling between body segments, 
consistent with the conservation of the normalized wavelength.

The experimental findings from Fig. 5 impose two constraints on a theory:
linear scaling between the velocities of propulsion and neuromuscular wave,
and the constancy of the slope. 
The mechanical part of our model enables us to gain insight about the
conditions under which the two empirical constraints are met.
Computation of a worm's propulsion velocity $v$ yields a following
relationship between $v$ and the neuromuscular velocity,
$\lambda\omega/2\pi$ (see, Appendix):

\begin{equation}
v = \gamma \frac{\lambda\omega}{2\pi},
\end{equation}\\
where the propeller efficiency coefficient $\gamma$ is given by

\begin{eqnarray*}
\gamma= \frac{\left[(c_{N}/c_{L})-1\right](2\pi A_{0}/\lambda)^{2}}  
{2 + (c_{N}/c_{L})(2\pi A_{0}/\lambda)^{2}}.   \nonumber 
\end{eqnarray*}\\
The result for $\gamma$ is essentially the same as that derived for flagellar
swimming on the basis of the Resistive Force theory 
(Childress, 1981).
The fact that $v$ is linearly related to $\omega$ follows from neglecting
worm's mass and associated inertial forces.
The coefficient $\gamma$ depends on the ratio of frictional coefficients
in the normal $c_{N}$ and longitudinal $c_{L}$ directions of the main
body axis, and on the ratio of the amplitude and wavelength. 
In the case relevant for worm's locomotion, $\gamma$ is positive and
in the range between 0 and 1. ($\gamma$ is negative for $c_{L} > c_{N}$
and in this case worm moves in the same direction as does neuromuscular
wave, which contradicts empirical observations and is not considered by us.)
The value of $\gamma$ approaches its maximal value 1, if any of the above 
ratios becomes large. Since the amplitude is weakly frequency dependent 
and can change by a factor of two among different mutants and species 
(Table 2), one could expect
that the coefficient $\gamma$ may be frequency dependent, and consequently,
the relationship between the velocities of propulsion and muscular wave
may have a non-linear character. However, the experimental data show
that these two quantities scale linearly with almost the same coefficient
$\gamma$, close to the optimal value 1, across wild-type and many mutant
{\it C. elegans}, and across related species (Fig. 5). Moreover, the ratio
$2\pi A_{0}/\lambda$ is of the order of 1 (Fig. 6, and Tables 1-3). 
This experimental linear scaling
can be reconciled with the formula (13), only if the ratio $c_{N}/c_{L}$
of the frictional coefficients is large, such that $\gamma$ becomes 
a frequency-independent constant approaching its maximal value 1.
Since the experimental data give $\gamma\approx 0.8$ for almost all cases, 
we can estimate from this the ratio of the frictional coefficients between 
worm's body and a substrate as $c_{N}/c_{L}\approx 9.0-14.0$. 
Surprisingly, this ratio is preserved even on substrates with increased 
firmness up to a large agar concentrations, which is evident from conserved
$\gamma$ in those cases (Table 3). This conservation presumably takes place 
because $c_{N}$ and $c_{L}$ change proportionally on these substrates such 
that their ratio remains constant.

A few mutants have slightly lower values of the coefficient $\gamma$. 
These are {\it sqt-1\/} mutants, defective in a cuticle collagen, 
and {\it cat-4\/} mutants, which are depleted in dopamine and serotonin 
and have weakened cuticle (Loer et al, 1999) (C. Loer, pers. comm.). 
They have the proportionality coefficient $\gamma$, around $0.5-0.6$, 
which corresponds to the significantly reduced ratio 
$c_{N}/c_{L}\approx 3.0-5.9$. This reduction is presumably due to 
some structural changes in the cuticle arrangement that impact its 
external condition, which in turn affect friction. However, 
even these exceptions still preserve the proportionality relation between
both velocities within their respective populations.

From Eq. (6), we determine the magnitude and spatio-temporal dependence of
the total bending moment $M(x,t)$. It depends non-linearly on the position
along the worm's length taking values zero at both endings and it exhibits
wave-like property (Fig. 7; and Appendix). This wave corresponds to
the neuro-muscular wave traveling from head ($x=0$) to tail ($x=L$).
Moreover, the magnitude of $M$ is proportional to the difference in the
neuro-muscular and propulsion velocities, i.e., it is proportional
to $1-\gamma$. For $\gamma$ close to its maximal value 1,
which is the case for {\it C. elegans}, its mutants, and related species, 
the magnitude of the total bending moment approaches zero (see Appendix). 
Thus, the efficient/optimal locomotion corresponds to the disappearance
of the total bending moment $M$ and the shear force $F$.
In this limit, the muscle bending moment $M_{m}$ exactly balances
body reaction represented by the visco-elastic bending moment $M_{e}$,
and worms do not slip.

Experimental data show negative correlations between the amplitude and
frequency of undulations (Tables 4 and 5). We seek to understand this 
inverse relationship in theoretical terms. To achieve this, we introduced 
explicitly simplified muscle dynamics into our model of undulatory 
movement and coupled it with the mechanics. This coupling allows
us to determine the dependence of the amplitude of undulations on the 
frequency and other parameters (see Appendix):

\begin{equation}
A_{0} = \frac{\lambda}{4\pi J}
\frac{\kappa fT}{\sqrt{[E^{2}+(\mu\omega)^{2}][(f+g)^{2}+\omega^{2}]}} 
\; + \; O(1-\gamma).
\end{equation}\\
Thus, the amplitude depends on several macroscopic and microscopic parameters:
the visco-elastic properties of the hydrostatic skeleton and cuticle 
characterized by $E$ and $\mu$, the neuromuscular synaptic transmission $T$, 
the rates characterizing contraction-relaxation cycle of muscle activity
$f$ and $g$, and the muscle stiffness $\kappa$. The correction $O$ in 
Eq. (14) is of the order of $(1-\gamma)$, which is small for the efficiency 
coefficient $\gamma$ is close to its upper limit 1. In general, formula 
(14) implies that the amplitude is a decreasing function of the frequency 
given that all other parameters are constant, which is the case for the worms 
with identical genomes. However, this dependence can be weak if characteristic
frequencies are smaller than parameters characterizing relaxation processes
$E/\mu$ (in cuticle/hydrostatic skeleton) and $f+g$ (in muscles).
This can explain weak negative correlations between the amplitude and 
frequency among worms representing the same genotype. Our experimental data 
are consistent with the decaying trend predicted 
by Eq. (14) both among worms representing the same species and among worms 
representing the same mutation (Fig. 8), although detailed fits are not 
possible due to large noise in the behavioral data.
This noise arises from the changes in the worm's direction of movement.
The fact that negative correlations are also present among different
mutants and among different species representing different genomes (Tab. 4)
indicates that the parameters $\kappa$, $f$, $g$, $T$, $E$, $\mu$, and $J$ 
in Eq. (14), do not vary dramatically in these worms. The possible 
variability of these parameters is not strong enough to reverse the decaying 
trend of the amplitude with the frequency.

From our theoretical analysis it follows that the amplitude in
Eq. (14) is proportional to the wavelength $\lambda$. Thus,
if frequency of undulations and other parameters in this equation do not 
change much, then the amplitude $A_{0}$ should scale linearly with $\lambda$.
Developmental locomotory data provide a good opportunity to verify this
scaling, since $\lambda$ changes by a factor of 4 during development.
Indeed, data from Fig. 6 confirm our theoretical result.

Experimental data from Table 2 indicate that the set of hyperactive mutants 
with affected synaptic signaling ({\it goa-1}, {\it egl-30}, {\it cat-2}, 
and partly {\it cat-4}) move faster and with higher frequencies than 
wild-type worms. These behavior is opposite to a more common behavior 
exhibited by most other neuronal mutations which cause locomotory slowdown. 
Such decreased velocity and frequency is also typical for mutants with 
structural changes in cuticle (BE109 and {\it sqt-1}) and muscle 
({\it unc-54}). Although our biomechanical model cannot explain why frequency
changes, it can explain the cross-dependencies between locomotory parameters.
We suspect that frequency increases in hyperactive neuronal mutants because 
they have elevated overall synaptic transmission that activates neurons and 
muscles more vigorously. On the other hand, for sluggish neuronal mutants 
frequency drops presumably because synaptic transmission is less effective.  
Muscle and cuticle structural mutants move with lower
frequency probably because they have modified neural activity via 
mechanosensory feedback. Because velocity and amplitude of
undulations depend on frequency, once the frequency is altered, they
change accordingly. In the case of velocity, there exists its linear
relationship with frequency (Fig. 5), and all mutants conform to 
that simple rule. The matter is more complex with the amplitude dependence 
on frequency, because several other factors are also involved (see Eq. 14).
This complexity likely explains why the general trend of negative correlations 
within a given genotype (Tables 4 and 5, and Fig. 8) does not always 
translate directly to such a dependence across different genotypes (Table 2).
For example, neuronal mutants in comparison to wild-type, despite having 
higher frequencies, do not have smaller amplitude. These mutants presumably 
have increased overall synaptic transmission $T$ at the neuromuscular 
junction (see Eq. (14)), which counterbalance and even overcomes the
effect associated with the frequency increase. Structural cuticle and muscle 
mutants have more predictable amplitude in comparison to the wild-type worms. 
Cuticle mutants have either comparable or increased amplitude, because their 
frequency is lower and their elasticity coefficient $E$ 
is slightly reduced, causing $A_{0}$ to increase in relation to wild-type 
(see Eq. 14). Similarly, muscle {\it unc-54}  mutants have in 
general increased amplitude because they have reduced frequency and 
because of the inverse dependence of $A_{0}$ on the crossbridge 
dissociation rate $g$. In {\it unc-54} mutants this rate is presumably 
decreased in comparison to wild-type (Moerman and Fire, 1997). Some
{\it unc-54} dominant-negative mutants have altered muscle structure
and are paralyzed (Bejsovec and Anderson, 1988, 1990). These mutants may
have severely reduced muscle stiffness coefficient $\kappa$, which leads
to $A_{0}\approx 0$ and consequently the propulsion velocity $v\approx 0$
(see Eq. (13)). Since the amplitude of undulations is proportional to the
synaptic transmisson $T$ at the neuromuscular junction and to the muscle
contraction rate $f$, so is the efficiency coefficient $\gamma$. However,
increase in either of these parameters will not necessarily lead to
a significant increase in $\gamma$ and $v$, because $\gamma$ is
already close to a saturation caused by a large ratio of $c_{N}/c_{L}$.
Increase in $\gamma$ and velocity would be much more pronounced in
environments in which $c_{N}$ and $c_{L}$ are comparable.

\vspace{2.5cm}

\noindent {\bf General discussion and summary}

We sought to systematize experimental behavioral 
data by providing a coherent model that could help us in understanding 
the main characteristics of the nematode movement and their 
interdependencies. In particular, our model points out the causes of 
the sinusoidal crawling and hints about its stability.

Our major experimental findings are: (i) the velocity of worm's propulsion
scales linearly with the velocity of the neuro-muscular wave with almost
optimal and highly conserved proportionality coefficient, (ii) the value 
of the normalized wavelength is conserved across a population of 
{\it C. elegans}, their mutants, and across related {\it Caenorhabditis} 
species, and (iii) the amplitude of undulations is inversely correlated with 
the frequency of the wave. These data provided enough information to
constrain our model and to derive some conclusions about parameters
controling the undulatory motion.

Our experimental data indicate that the frequency of undulations
varies between different {\it Caenorhabditis} species, between different 
mutants of {\it C. elegans}, and even between different worms representing 
the same genotype. This observation suggests that the presumed
oscillatory activity of motor neurons can be easily modulated, which 
can be explained by noting that nerve ring input and presumably 
mechanosensory receptors are involved in modulating ocillatory activity
of motor neurons. Since both of these contributions can depend on the
inner state of an animal, it is not surprising that the frequency 
of undulations can be altered for the worms with the same genotype. 
For example, we expect that mutations to genes involved in inter-neuronal 
transmission ({\it goa-1} and {\it egl-30} encoding proteins that regulate 
synaptic transmission, {\it cat-2} involved in synthesis of dopamine, and 
{\it cat-4\/} involved in synthesis of dopamine and serotonin) affect the 
input coming to motor neurons and consequently modulate the frequency of
their activity. However, our biomechanical model cannot explain precisely
why frequency changes under different conditions. This would require
an approach that explicitly takes neural dynamics aspects into account.
Nevertheless, our model can explain relationships of other locomotory 
parameters to the frequency.

Nematode worms do not possess limbs and therefore have to use other
strategies for efficient locomotion. These worms move primarily because 
of two main factors: generation of the neuro-muscular wave, and the presence 
of oscillations in local units between activities of dorsal and ventral 
parts of the body. The second factor, characterized by the frequency,
is the driving force for locomotion, since it enables the wave
to acquire speed (equal to $\lambda\omega/2\pi$) and consequently to
propagate along the body. The traveling wave is crucial for the movement,
since if it does not propagate, then the worm does not move (non-propagating
wave has $\lambda\omega = 0$ and, from Eq. (13), $v= 0$). In the case
when there are no oscillations, the neuromuscular wave can still exist
(provided stretch receptors are active) but only as a standing wave that
bends the body, which however is not enough to generate worm's movement.
Indeed, {\it C. elegans} often pause, and maintain their body posture,
then resume movement.

The wave can propagate in either direction of the body and this is
strictly related to the direction of worm's motion. If the wave propagates
backward, then the worm's muscles exert force on the ground that is directed 
toward the tail. The mechanics principle of action and 
contraction (the environment reacts with the opposite force on the 
worm's body) is responsible for the worm's forward motion.
If the neuro-muscular wave propagates forward along the body, then
the directions of the forces are reversed, and consequently the worm
moves backward. From neurophysiological data it is evident that the
direction of the wave is correlated with the direction of dendritic
processes of motor neurons in the forward and backward motion neural
circuits (Chalfie et al, 1985; White et al, 1986).
This observation was the motivation for Byerly and Russell (cited in White
et al, 1986) to propose that dendrites may contain stretch receptors
that are used in wave generation.

Oscillations in local units create anti-phase activity between dorsal
and ventral parts of the body, which in turn, is responsible for a 
nonzero amplitude of undulations. The nonzero value of this amplitude 
is necessary for transforming part of
the neuro-muscular wave velocity into the velocity of propulsion. This
can be seen from Eq. (13), which relates the two velocities and contains
the explicit dependence of the propeller efficiency coefficient $\gamma$
on the undulatory amplitude. The greater the amplitude of undulations,
the greater the efficiency of movement (i.e., the propulsion velocity
is greater for a given velocity of the neuro-muscular wave).
However, this increasing trend has its limit, since amplitude and frequency
are negatively correlated (see Tables 4 and 5). Thus, at higher frequencies 
the amplitude decreases, and the movement efficiency would decrease. 
The fact that the slopes of the regression lines and some 
data points in Fig. 4 are close to the maximal allowed value for $\gamma$
suggests that {\it Caenorhabditis} worms are evolutionarily adapted 
to maximize this coefficient. However, based on our estimation of parameters,
they do it not by increasing the amplitude but by increasing the ratio 
of the frictional coefficients $c_{N}/c_{L}$. This presumably takes place 
because of the longitudinal arrangement of structural elements in the cuticle 
that influences its external condition such that it enables worms to move 
much easier longitudinally than normally to the main body axis. 
This choice of the 
optimizing parameter is more beneficial, since it potentially maximizes 
$\gamma$ in different environments regardless of the values of other
parameters.

Although all mutations examined by us change the average frequency of 
undulations, they do not have disastrous effects on movement. 
(There are of course mutations that abolish movement; Brenner, 1974.)
This observation suggests that the stability of the neuro-muscular 
traveling wave is robust against perturbations, and our model can provide 
an explanation for it. To destroy the traveling wave, one would have 
to either eliminate oscillations or to alter the long-range coupling 
between worm's body segments which is presumably mediated by long neuronal
processes. Our genetic perturbations do not change the anatomical structure 
of the nervous system, instead, some of them change local chemical signaling 
between neurons. This, however, only modulates the wave speed by changing 
its frequency, and does not influence the wave stability.

We found that the amplitude and frequency of undulations are negatively
correlated. Especially, the correlation between instantaneous values
of these parameters is important, since it captures sudden changes
in worm's movement. All wild-type species and most {\it C. elegans}
mutants show robust negative instantaneous correlations with very small
p-values ($p < 10^{-4}$). However, for some {\it unc-54} mutants p-values 
are large, which suggests that correlations are not statistically 
significant in those cases. This can be explained by the formula in Eq. (14). 
{\it unc-54} worms move much slower than the wild-type and their
frequency is small. Equation (14) implies that for sufficiently small 
frequencies, amplitude of undulations becomes practically frequency
independent. Thus, any sudden change in frequency would have almost no effect
on amplitude, which explains why these two parameters are poorly correlated
in these cases.

The mechanical part of our model assumes that components of the resistive 
force acting on nematodes are proportional to the corresponding components 
of the velocity of motion. This assumption, called the Resistive Force 
Theory, is borrowed from fluid dynamics models describing swimming 
(Gray and Hancock, 1955; Lighthill, 1976). In our experiments, the worms 
do not swim. Instead they crawl through a thin layer of bacteria and therefore 
the law of the Resistive Force Theory may be valid only approximately.
It would be interesting to investigate a more realistic resistance law 
applicable to our locomotory system, although experimentally it may be
hard to achieve due to microscopic sizes of nematodes.

In summary, we presented an integrated approach to studying the 
undulatory locomotion of nematode worms. Since physiological approaches
are difficult in {\it C. elegans}, our experimental approach consisted
of genetic perturbations to parameters controling the movement. 
By comparing different mutants and {\it Caenorhabditis} species we were 
able to construct a biomechanical theoretical model that provides insight 
about different factors involved. Our results reveal that although 
different mechanical parameters characterizing the undulatory locomotion 
change as we apply perturbations, their interdependencies are robust and 
do not fall apart easily. In particular, we have found optimality of the 
motion reflected in the value of the efficiency coefficient close to its 
maximal value. This optimality corresponds to almost exact balance between
the muscle bending moment and the visco-elastic body reaction moment.
Conserved relationships between locomotory parameters suggest robust 
cellular and molecular controlling mechanisms that can be fleshed out with 
the extensive knowledge of details of the {\it C. elegans} nervous system.

\vspace{3cm}

\noindent {\bf Appendix}

In this Appendix we provide more details on some derivations.
Solution of Eq. (6) in the main text yield the following spatio-temporal 
dependence of the total bending moment $M(x,t)$:

\begin{eqnarray}
M(x,t)= \frac{c_{N}A_{0}\lambda^{2}\omega}{4\pi^{2}}(1-\gamma) \left[
\sin(\omega t -kx) - \sin(\omega t) 
+ kx\cos(\omega t) \right.   \nonumber  \\
  - (kx^{2}/L) \left[\cos(\omega t - kL) 
+ 2 \cos(\omega t) + (3/kL) (\sin(\omega t - kL) 
- \sin(\omega t) ) \right]   \nonumber  \\ 
\left.  + (kx^{3}/L^{2})\left[\cos(\omega t - kL) 
+ \cos(\omega t) + (2/kL) (\sin(\omega t - kL) 
- \sin(\omega t) ) \right] \right],
\end{eqnarray}\\
where $k= 2\pi/\lambda$. $M(x,t)$ is composed of the traveling wave given by 
$\sin(\omega t - kx)$ and the standing wave (the rest of the 
terms). The latter originates from the recoil correction $\delta h(x,t)$.
It is important to note that the total bending moment is proportional to 
$1-\gamma$, and thus for the efficiency coefficient $\gamma$ approaching its 
upper limit 1, the moment $M$ is zero.

The explicit form of the visco-elastic bending moment $M_{e}(x,t)$ 
obtained by inserting Eq. (7) into Eq. (8) is:

\begin{eqnarray}
M_{e}(x,t)=  \frac{4\pi^{2} A_{0}J}{\lambda^{2}}
\left[ E\cos(\omega t - kx)  - 
\mu\omega\sin(\omega t - kx) \right].
\end{eqnarray}\\
This moment as well as the total bending moment are both proportional
on the amplitude $A_{0}$ of undulations. Its space averaged form is:

\begin{eqnarray}
\overline{M_{e}}(t)=  \frac{4\pi A_{0}J}{\lambda L}
\left( E \sin\omega t  + \mu\omega\ \cos\omega t  \right),
\end{eqnarray}\\
where we used experimental fact that $\lambda\approx 2L/3$.

The spatially averaged muscle bending moment $\overline{M_{m}}$ is 
determined from solving Eqs. (10)-(12) in the main text. It has
the form:

\begin{eqnarray}
\overline{M_{m}}(t)= \frac{\kappa fT}{L[(f+g)^{2} + \omega^{2}]}
\left[ (f+g)\cos(\omega t + \xi) + \omega\sin(\omega t + \xi) \right].
\end{eqnarray}

The dependence of the amplitude of undulations $A_{0}$ on the frequency
$\omega$ can be determined from the above equations using the facts
that $M= M_{e} + M_{m}$, and $M \approx 0$ in the limit 
$\gamma \approx 1$, which is justified experimentally. 
Thus, in this limit we have a perfect balance of $M_{m}$ and $M_{e}$, i.e., 
$\overline{M_{m}} \approx -\overline{M_{e}}$.
Because both moments $\overline{M_{m}}$ and $\overline{M_{e}}$ oscillate 
in time, the next step is to
take squares of both sides of this equation and then to perform temporal
averaging over one period of oscillation, i.e.,

\begin{equation}
\langle\overline{M_{m}}^{2}\rangle=  \langle\overline{M_{e}}^{2}\rangle  
\;\; + \;\; O(1-\gamma).
\end{equation}\\
The averaging is done using the facts that $\langle\sin^{2}\omega t\rangle
= \langle\cos^{2}\omega t\rangle = 1/2$ and
$\langle\sin\omega t \;\cos\omega t\rangle = 0$. 
As a result of this procedure, we obtain Eq. (14) in the main text.

The propulsion velocity $v$ is determined from Eq. (9) using Eqs. (7) and
(15) for the lateral displacement $h$ and the total bending moment $M$, 
respectively. After tedious algebra we obtain:

\begin{equation}
\int_{0}^{L} dx \langle \frac{\partial M}{\partial x} 
\frac{\partial^{2}h}{\partial x^{2}}\rangle = 
\frac{1}{2}c_{N}A_{0}^{2}kL(\omega - vk)
\left[ 1 + O\left(1/(kL)^{2}\right)\right],
\end{equation}\\
where the main contribution comes from the traveling wave part of $M$,
and

\begin{equation}
\int_{0}^{L} dx \langle \frac{\partial h}{\partial t} 
\frac{\partial h}{\partial x} \rangle =
 - \frac{1}{2}A_{0}^{2}\omega kL
\left[ 1 + O\left(1/(kL)\right)\right],
\end{equation}\\
where again the main contribution comes from the traveling wave part
of $h(x,t)$. The correction to the velocity from the recoil term is
small, of the order of $1/(kL)\approx 0.1$.
In general, $v$ is approaching the velocity of the neuromuscular wave, 
$\lambda\omega/2\pi$, if $c_{N}/c_{L}\mapsto \infty$.

\vspace{4.5cm}

\noindent{\bf Acknowledgments}

We thank J. Kramer for {\it sqt-1\/} and BE109 mutants and unpublished
data, and Charles Brokaw and Erich Schwarz for comments on a draft of
this manuscript. Some strains were obtained from the Caenorhabditis
Genetics Center. 
The work was supported by the Sloan-Swartz fellowship at 
Caltech (J.K.), by the Howard Hughes Medical Institute, with which
P.W.S. is an Investigator, and by DARPA. A.S. is an HHMI pre-doctoral
fellow.

\vspace{2.5cm}

\noindent{\bf References} \\
Alexander RMcN, Goldspink G (1977). 
Mechanics and Energetics of Animal Locomotion. London:
Chapman and Hall. \\
Bargmann CI (1998). Neurobiology of the {\it Caenorhabditis
elegans} genome. Science 282: 2028-2033. \\
Bejsovec AM, Anderson P (1988). Myosin heavy chain
mutations that disrupt Caenorhabditis elegans thick filament assembly.
Genes and Develop. 2: 1307-1317; and (1990) 
Functions of the myosin ATP and actin binding sites are required for 
C. elegans thick filament assembly. Cell 60: 133-140. \\
Brenner S (1974). The genetics of Caenorhabditis elegans.
Genetics  77: 71-94. \\
Cang J, Friesen WO (2002). Model for intersegmental 
coordination of leech swimming: Central and sensory mechanisms.
J. Neurophysiol. 87: 2760-2769. \\
Chalfie M et al. (1985). The neural circuit for touch 
sensitivity in {\it Caenorhabditis elegans.} J. Neurosci. 5: 956-964. \\
Chalfie M, White J (1988). {\it The Nervous System.} 
In: Wood WB, editor. The Nematode Caenorhabditis elegans.
Cold Spring Harbor: Cold Spring Harbor Laboratory Press, 
pp. 337-391. \\
Cheng JY, Pedley TJ, Altringham JD (1998).
A continuous dynamic beam model for swimming fish. Phil. Trans. 
R. Soc. Lond. B 353: 981-997. \\
Childress S (1981). Mechanics of swimming and flying. Cambridge: Cambridge
Univ. Press. \\
Cohen AH et al. (1992). Modeling of intersegmental 
coordination in the lamprey central pattern generator for locomotion. 
Trends Neurosci. 15: 434-438.  \\
Cortez R et al. (2004). Simulation of swimming organisms: Coupling
internal mechanics with external fluid dynamics.
Comp in Sci and Eng, IEEE 6: 38-45. \\
Cronin CJ et al. (2005). An automated system for measuring parameters
of nematode sinusoidal movement. BMC Genet. 6: 5. \\
Davis RE, Stretton AOW (1989). Signaling properties
of {\it Ascaris} motorneurons: graded active responses, graded synaptic
transmission and tonic transmitter release. J. Neurosci. 9: 
415-425. \\
de Bono M, Maricq AV (2005). Neuronal substrates of complex behaviors
in {\it C. elegans}. Annu. Rev. Neurosci. 28: 451-501. \\
Francis MM, Mellem JE, Maricq AV (2003). Bridging
the gap between genes and behavior: recent advances in the 
electrophysiological analysis of neural function in {\it Caenorhabditis
elegans}. Trends Neurosci. 26: 90-99. \\
Friesen WO, Cang J (2001). Sensory and central mechanisms
control intersegmental coordination. Curr. Opin. Neurobiol. 11:
678-683. \\
Gray J, Hancock GJ (1955). The propulsion of sea-urchin spermatozoa.
J. Exp. Biol. 32: 802-814. \\
Gray J (1964). Animal Locomotion. London: Weidenfeld and Nicolson. \\
Gray J, Lissmann HW (1964). The locomotion of nematodes.
J. Exp. Biol. 41: 135-154. \\
Gutkin BS, Ermentrout GB, Reyes AD (2005). Phase-response curves give the
responses of neurons to transient inputs. J. Neurophysiol. 94: 1623-1635.\\
Hirose S (1993). Biologically inspired robots: snake-like locomotors
and manipulators. New York: Oxford Univ. Press. \\
Huxley AF (1957). Muscle structure and theories of contraction.
Prog. Biophys. Biophys. Chem.  7: 255-318. \\
Ijspeert AJ (2001). A connectionist central pattern generator for the aquatic
and terrestrial gaits of a simulated salamander. Biol. Cybern. 84: 331-348.\\
Kapatos G et al. (1999). GTP cyclohydrolase I feedback 
regulatory protein is expressed in serotonin neurons and regulates 
tetrahydrobiopterin biosynthesis. J. Neurochem. 72: 669-675. \\
Kiontke K, Hironaka M, Sudhaus W (2002). Description
of Caenorhabditis japonica n. sp (Nematoda: Rhabditida) associated with
the burrower bug Parastrachia japonensis (Heteropter: Cydnidae) in Japan.
Nematology  4: 933-941. \\
Kopell N, Ermentrout GB (1988). Coupled oscillators and
the design of central pattern generators. Math. Biosci.
90: 87-109. \\
Kramer JM et al. (1988). The sqt-1 gene of C. elegans encodes
a collagen critical for organismal morphogenesis. Cell  55: 
555-565. \\
Lighthill MJ (1960). Note on the swimming of slender fish.
J. Fluid Mech. 9: 305-317. \\
Lighthill MJ (1976). Flagellar hydrodynamics.
SIAM Review 18: 161-230. \\
Lints R, Emmons SW (1999). Patterning of dopaminergic 
neurotransmitter identity among Caenorhabditis elegans ray sensory
neurons by a TGF$\beta$ family signaling pathway and a Hox gene.
Develop. 126: 5819-5831. \\
Lockery SR, Sejnowski TJ (1993). The computational leech. Trends Neurosci.
16: 283-290. \\
Loer CM, Davidson B, Mckerrow J (1999). A phenylalanine
hydroxylase gene from the nematode C. elegans is expressed in the hypodermis.
J. Neurogenet. 13: 157-180. \\
Maduzia LL, et al. (2002). lon-1 regulates Caenorhabditis 
elegans body size downstream of the dbl-1 TGF beta signaling pathway.
Develop. Biol. 246: 418-428.  \\
Marder E, Calabrese RL (1996). Principles of rhythmic
motor pattern generation.  Physiol. Rev. 76: 687-717. \\
Mendel JE et al. (1995). Participation of the protein Go
in multiple aspects of behavior in C. elegans.  Science 267: 
1652-1655. \\
Moerman DG, Fire A (1997). {\it Muscle: Structure, Function,
and Development. } In: Riddle DL et al. editors. C. elegans II. 
Cold Spring Harbor: Cold Cold Spring Harbor Laboratory Press, 
pp. 417-470. \\
Moghal N et al. (2003). Modulation of EGF receptor-mediated
vulva development by the heterotrimeric G protein G-alpha q and excitable
cells in C. elegans. Develop. 130: 4553-4566. \\
Niebur E, Erdos P (1991). Theory of the locomotion of
nematodes. Biophys. J. 60: 1132-1146. \\
Pedley TJ, Hill SJ (1999). Large-amplitude undulatory fish swimming:
fluid mechanics coupled to internal mechanics. J. Exp. Biol. 202: 3431-3438. \\
Segalat L, Elkes DA, Kaplan JM (1995). Modulation
of serotonin-controlled behaviors by Go in Caenorhabditis elegans.
Science  267: 1648-1651. \\
Skinner FK, Mulloney B (1998). Intersegmental coordination
in invertebrates and vertebrates. Curr. Opin. Neurobiol.  8:
725-732. \\
Sulston JE, Dew M, Brenner S (1975). Dopaminergic
neurons in the nematode C. elegans. J. Comp. Neurology 163:
215-226. \\
White JG, Southgate E, Thomson JN, Brenner S (1986).
The structure of the nervous system of the nematode 
{\it Caenorhabditis elegans.}  Phil. Trans. R. Soc. Lond. B 314: 
1-340. \\
Williams T (1998). Predicting force generation by lamprey muscle
during applied sinusoidal movement using a simple dynamic model.
J. Exp. Biol. 201: 869-875. \\
Wu TY (1971). Hydrodynamics of swimming fishes and cetaceans.
Adv. Appl. Math. 11: 1-63.

\newpage

\begin{center}

\begin{tabular}{|l l l l l l l l|}
\multicolumn{8}{l}
{ Table 1.}\\
\multicolumn{8}{l} 
{Forward locomotory data for wild-type {\it C. elegans} during different
developmental stages.} \\

  \hline \hline

 Development & N &  $v$ [mm/s]  & $\omega/2\pi$ [Hz] & $A_{0}$ [mm] 
 & $\lambda$ [mm]  &  $\lambda/L$  & $\gamma$  \\
 
 stage         & & & & & & & \\
    
\hline\hline

 L1     & 4 & 0.04$\pm$0.02 & 0.26$\pm$0.12 & 0.04$\pm$0.01 
  & 0.18$\pm$0.02   & 0.71$\pm$0.05 &  0.93$\pm$0.61 \\

 L2     & 4 & 0.08$\pm$0.01 & 0.34$\pm$0.06 & 0.06$\pm$0.00 
  & 0.27$\pm$0.02  &  0.59$\pm$0.03 & 0.82$\pm$0.18 \\
  
 L3     & 5 &  0.08$\pm$0.02 & 0.26$\pm$0.06 & 0.08$\pm$0.01
  & 0.36$\pm$0.01 &  0.63$\pm$0.01  &   0.84$\pm$0.28  \\

 L4     & 5  & 0.15$\pm$0.05 & 0.31$\pm$0.12 & 0.11$\pm$0.01
  & 0.54$\pm$0.01 &  0.66$\pm$0.02  &   0.89$\pm$0.45 \\

 Adult  &  5   & 0.12$\pm$0.03 & 0.20$\pm$0.07 & 0.14$\pm$0.03
  & 0.70$\pm$0.03 &  0.65$\pm$0.02  &   0.86$\pm$0.37  \\

\hline \hline

\end{tabular}

\end{center}

\noindent N is the number of worms used in every developmental stage.
Parameter $\gamma$ is defined as $\gamma= v/(\lambda\omega/2\pi)$.

\newpage

\begin{center}

\begin{tabular}{|l l l l l l l l|}
\multicolumn{8}{l}
{ Table 2.}\\
\multicolumn{8}{l} 
{Forward locomotory data for adult wild-type (WT) {\it C. elegans}, 
its mutants,} \\
\multicolumn{8}{l} 
{and related species.} \\
  \hline \hline

 Genotype  & N  & $v$ [mm/s]  & $\omega/2\pi$ [Hz] & $A_{0}$ [mm] 
   & $\lambda$ [mm]   &  $\lambda/L$  & $\gamma$  \\
    
\hline\hline

{\it C. elegans} WT  & 58  & 0.17$\pm$0.04 & 0.36$\pm$0.08 & 0.09$\pm$0.01
  & 0.59$\pm$0.04 &  0.62$\pm$0.02  &   0.79$\pm$0.26  \\

   Mutants:  & & & & & &  &  \\ 
  
 $\;$  {\it cat-2(e1112)}  & 13  & 0.24$\pm$0.04 & 0.44$\pm$0.06 
  & 0.11$\pm$0.01 & 0.67$\pm$0.04 &  0.62$\pm$0.02  &   0.82$\pm$0.18  \\

 $\;$ {\it cat-4(e1141)} & 10  & 0.12$\pm$0.03 & 0.43$\pm$0.06 
  & 0.12$\pm$0.02 & 0.57$\pm$0.02 &  0.62$\pm$0.02  &   0.48$\pm$0.13 \\

 $\;$ {\it egl-30(tg26)}  & 16  & 0.19$\pm$0.03 & 0.57$\pm$0.08 
  & 0.10$\pm$0.00 & 0.41$\pm$0.02 &  0.54$\pm$0.02  &   0.81$\pm$0.17  \\

 $\;$  {\it goa-1(n1134)}  & 12  & 0.22$\pm$0.05 & 0.53$\pm$0.10 
 & 0.09$\pm$0.01  & 0.54$\pm$0.04 &  0.63$\pm$0.02  &   0.76$\pm$0.23 \\

 $\;$ {\it goa-1(sy192)}  &  11  & 0.24$\pm$0.03 & 0.58$\pm$0.06 
 & 0.11$\pm$0.01  & 0.53$\pm$0.02 &  0.64$\pm$0.01  &   0.76$\pm$0.12 \\

 $\;$ {\it lon-1(e185)}  & 5  & 0.21$\pm$0.03 & 0.40$\pm$0.03 
  & 0.13$\pm$0.01 & 0.67$\pm$0.03 &  0.62$\pm$0.02  &   0.76$\pm$0.12 \\

 $\;$ {\it lon-1(e185);}  & & & & & & &   \\
 $\;\;\;\;$  {\it lon-2(e678)}  & 5  & 0.15$\pm$0.03 & 0.33$\pm$0.08 
  & 0.16$\pm$0.02 & 0.60$\pm$0.05 &  0.56$\pm$0.06  &   0.75$\pm$0.24  \\

  $\;$   BE109   & 9  & 0.07$\pm$0.03 & 0.18$\pm$0.06 
  & 0.11$\pm$0.01 & 0.43$\pm$0.04 &  0.57$\pm$0.04  &   0.91$\pm$0.48 \\

  $\;$ {\it sqt-1(sc101)}  & 13  & 0.10$\pm$0.03 & 0.26$\pm$0.07 
  & 0.12$\pm$0.01 & 0.69$\pm$0.04 &  0.63$\pm$0.03  &   0.55$\pm$0.22 \\

  $\;$ {\it sqt-1(sc103)}   & 5  & 0.09$\pm$0.03 & 0.27$\pm$0.10 
  & 0.09$\pm$0.02 & 0.53$\pm$0.04 &  0.62$\pm$0.04  &   0.62$\pm$0.31 \\

 $\;$  {\it unc-54(st130)}  & 5  & 0.02$\pm$0.01 & 0.05$\pm$0.00 
  & 0.11$\pm$0.02 & 0.55$\pm$0.04 &  0.69$\pm$0.03  &   0.73$\pm$0.36 \\

 $\;$ {\it unc-54(st132)}   & 5  & 0.08$\pm$0.01 & 0.15$\pm$0.03 
  & 0.10$\pm$0.01 & 0.59$\pm$0.02 &  0.65$\pm$0.03  &   0.90$\pm$0.21 \\

 $\;$  {\it unc-54(st134)}   & 5  & 0.05$\pm$0.01 & 0.08$\pm$0.01 
  & 0.10$\pm$0.01 & 0.56$\pm$0.03 &  0.65$\pm$0.03  &   1.16$\pm$0.28 \\

 $\;$  {\it unc-54(st135)}   & 5  & 0.02$\pm$0.01 & 0.05$\pm$0.01 
  & 0.16$\pm$0.02 & 0.57$\pm$0.01 &  0.62$\pm$0.02  &   0.60$\pm$0.32 \\

 $\;$  {\it unc-54(s95)}   & 5  & 0.04$\pm$0.01 & 0.07$\pm$0.01 
  & 0.12$\pm$0.03 & 0.63$\pm$0.04 &  0.66$\pm$0.02  &   1.00$\pm$0.29 \\

 $\;$ {\it unc-54(s74)}    & 5  & 0.04$\pm$0.01 & 0.08$\pm$0.02 
  & 0.13$\pm$0.01 & 0.48$\pm$0.02 &  0.58$\pm$0.02  &   0.97$\pm$0.34 \\

\hline

Species:  & & & & & &  &   \\ 

 $\;$   {\it C. briggsae}  &  5  & 0.15$\pm$0.06 & 0.30$\pm$0.12 
 & 0.09$\pm$0.02  & 0.56$\pm$0.02 &  0.59$\pm$0.02  &   0.91$\pm$0.51  \\

 $\;$  {\it C. japonica}  &  5  & 0.08$\pm$0.03 & 0.15$\pm$0.04 
 & 0.15$\pm$0.02  & 0.69$\pm$0.04 &  0.68$\pm$0.03  &   0.71$\pm$0.32 \\

 $\;$  {\it C. remanei}  &  5  & 0.25$\pm$0.06 & 0.41$\pm$0.09 
 & 0.15$\pm$0.02  & 0.67$\pm$0.04 &  0.64$\pm$0.03  &   0.93$\pm$0.30 \\

 $\;$  PS1010 {\it Caen.}  &  4  & 0.19$\pm$0.01 & 0.39$\pm$0.04 
 & 0.11$\pm$0.01  & 0.54$\pm$0.04 &  0.68$\pm$0.03  &   0.93$\pm$0.12 \\

 $\;$  CB5161 {\it Caen.}  &  5  & 0.14$\pm$0.03 & 0.29$\pm$0.05 
 & 0.13$\pm$0.02  & 0.67$\pm$0.06 &  0.62$\pm$0.02  &   0.71$\pm$0.20 \\
   
\hline \hline

\end{tabular}

\end{center}

\newpage

\begin{center}

\begin{tabular}{|l l l l l l l|}
\multicolumn{7}{l}
{ Table 3.}\\
\multicolumn{7}{l} 
{Forward locomotory data for adult wild-type {\it C. elegans} on different 
substrates. } \\

  \hline \hline

 Agar percentage & $v$ [mm/s]  & $\omega/2\pi$ [Hz] & $A_{0}$ [mm] 
                          & $\lambda$ [mm]   &  $\lambda/L$  & $\gamma$  \\
    
\hline\hline

 2 $\%$  (N=18) & 0.21$\pm$0.04 & 0.39$\pm$0.08 & 0.13$\pm$0.02 
  & 0.72$\pm$0.03   & 0.66$\pm$0.02 &  0.75$\pm$0.21 \\

 4 $\%$ (N=18) & 0.17$\pm$0.04 & 0.34$\pm$0.06 & 0.11$\pm$0.02 
  & 0.64$\pm$0.05  &  0.62$\pm$0.04 &  0.77$\pm$0.23 \\

 6 $\%$ (N=17) & 0.15$\pm$0.04 & 0.34$\pm$0.06 & 0.12$\pm$0.03 
  & 0.62$\pm$0.07  &  0.62$\pm$0.03 &  0.71$\pm$0.24 \\

 8 $\%$ (N=17) & 0.10$\pm$0.02 & 0.27$\pm$0.05 & 0.13$\pm$0.02 
  & 0.61$\pm$0.05  &  0.62$\pm$0.04 &  0.63$\pm$0.17 \\

\hline \hline

\end{tabular}

\end{center}

\newpage

\begin{center}

\begin{tabular}{|l l l|}
\multicolumn{3}{l}
{ Table 4.}\\
\multicolumn{3}{l} 
{ Correlation between the amplitude $A_{0}$ and frequency $\omega$ of 
for the forward} \\
\multicolumn{3}{l} 
{motion of wild-type {\it C. elegans}, its mutants and related species.} \\
  \hline \hline

 Genotype   &  Correlation coefficient:  &  Correlation coefficient:  \\
     &  instantaneous $A_{0}$ and $\omega$  &  average $A_{0}$ and $\omega$  \\
\hline\hline

{\it C. elegans} Wild-type  &        &     \\
 $\;\;$  L1  stage   &  -0.417 (0.000; 3268)   &   -0.812 (0.188; 4) \\

 $\;\;$  L2  stage   &  -0.169 (0.000; 2410)   &   -0.904 (0.096; 4) \\
  
 $\;\;$  L3  stage   &  -0.290 (0.000; 4278)   &    -0.770 (0.128; 5) \\

 $\;\;$  L4  stage   &  -0.350 (0.000; 4308)   &    -0.966 (0.008; 5) \\

 $\;\;$  Adult       &  -0.251 (0.000; 68302)  &    -0.505 (0.000; 58)   \\

   Mutants:  &   &  \\ 
  
 $\;\;$  {\it cat-2(e1112)}   &  -0.286 (0.000; 16034)   & -0.714 (0.006; 13)  \\

 $\;\;$ {\it cat-4(e1141)}   &  -0.149 (0.000; 7872)   & -0.444 (0.199; 10)   \\

 $\;\;$ {\it egl-30(tg26)}   &  -0.206 (0.000; 10431)  & -0.228 (0.500; 11)   \\

 $\;\;$  {\it goa-1(n1134)}   &  -0.300 (0.000; 12922)   & -0.537 (0.072; 12)  \\

 $\;\;$ {\it goa-1(sy192)}   &  -0.244 (0.000; 11017)  & -0.391 (0.235; 11)   \\

 $\;\;$ {\it lon-1(e185)}   &  -0.214 (0.000; 17296)  & -0.799 (0.105; 5)  \\

 $\;\;$ {\it lon-1(e185); lon-2(e678)}  &  -0.104 (0.000; 10670)  &  0.266 
(0.665; 5)  \\

  $\;\;$   BE109          &   -0.168 (0.000; 10000)  & -0.256 (0.505; 9)  \\

  $\;\;$ {\it sqt-1(sc101)}    &  -0.189 (0.000; 14038)   & -0.442 
(0.131; 13)  \\

  $\;\;$ {\it sqt-1(sc103)}    &  -0.382 (0.000; 5551)   &  -0.963 
(0.008; 5) \\

 $\;\;$  {\it unc-54(st130)}    &   -0.002 (0.918; 3785)  &  -0.935 
(0.020; 5)  \\

 $\;\;$ {\it unc-54(st132)}    &  -0.237 (0.000; 4968)   &  -0.027 
(0.966; 5)  \\

 $\;\;$  {\it unc-54(st134)}    &  -0.024 (0.083; 5457)  &  -0.629 
(0.256; 5)  \\

 $\;\;$  {\it unc-54(st135)}    &  0.0068 (0.675; 3824)  &  -0.184 
(0.767; 5)  \\

 $\;\;$  {\it unc-54(s95)}   &   -0.026 (0.058; 5506)  & -0.955 (0.011; 5) \\

 $\;\;$ {\it unc-54(s74)}    &  -0.060 (0.001; 2965)  &  -0.915 
(0.029; 5) \\ \hline

 $\;\;$  mutants cross-correlation  &      &   -0.293 (0.247; 17)  \\  \hline

Species:  &   &   \\ 

 $\;\;$   {\it C. briggsae}  &  -0.432 (0.000; 7061)   &  -0.974 
(0.005; 5)  \\

 $\;\;$  {\it C. japonica}  &  -0.255 (0.000; 4659)    &  -0.790 
(0.112; 5)  \\ 

 $\;\;$  {\it C. remanei}  &  -0.380 (0.000; 4955)   &  -0.335 (0.582; 5)  \\

 $\;\;$  PS1010 {\it Caenorhabditis}   &  -0.497 (0.000; 5049)  & -0.831 
(0.169; 4)  \\

 $\;\;$  CB5161 {\it Caenorhabditis}   &  -0.266 (0.000; 5172)  &  0.236   
(0.703; 5)  \\ \hline
   
 $\;\;$  species cross-correlation  &      &    -0.329 (0.262; 6)  \\  
\hline \hline

\end{tabular}

\end{center}

For each correlation coefficient we show in the bracket corresponding 

p-value and the number of data pair points, respectively.

\newpage

\begin{center}

\begin{tabular}{|l l l|}
\multicolumn{3}{l}
{ Table 5.}\\
\multicolumn{3}{l} 
{ Correlation between the amplitude $A_{0}$ and frequency $\omega$ of 
for the forward} \\
\multicolumn{3}{l} 
{motion of adult wild-type {\it C. elegans} on different substrates.} \\
  \hline \hline

 Agar percentage   &  Correlation coefficient:  &  Correlation coefficient:  \\
     &  instantaneous $A_{0}$ and $\omega$  &  average $A_{0}$ and $\omega$  \\
\hline\hline

 $2 \%$    &  -0.325 (0.000; 17378)   &   -0.246 (0.325; 18) \\

 $4 \%$   &  -0.213 (0.000; 19421)   &   -0.099 (0.697; 18) \\
  
 $6 \%$   &  -0.183 (0.000; 17828)   &    -0.270 (0.294; 17) \\

 $8 \%$   &  -0.155 (0.000; 17950)   &    -0.383 (0.129; 17) \\

\hline \hline

\end{tabular}

\end{center}

For each correlation coefficient we show in the bracket corresponding 

p-value and the number of data pair points, respectively.

\newpage

{\bf \large Figure Captions}

Fig. 1\\
Photographs of typical shapes of worms performing undulatory locomotion 
recorded in our experiments. An extreme example is {\it unc-54(st135)}
worm, which moves very slowly, and with low frequency but has a large
amplitude of the wave (panel d). The white midline with
dots on it was used as a reference frame for quantifying the movement.

\vspace{0.3cm}

Fig. 2\\
The system controling locomotory output in {\it Caenorhabditis}.
Nerve ring neurons activate motor neurons, which in turn activate 
body-wall muscles. Muscle activity causes the worm to move and the movement
is modulated by hydrostatic skeleton and cuticle, and by mechanosensory
feedback. The latter component provides stability for the undulations. 
The lower panel shows basic components of the neural structure for forward 
locomotion: the nerve ring, ventral nerve cord (dashed line), and
excitatory B motor neurons. Neuronal processes in B motor neurons 
(only 3 shown) are elongated and it is hypothesized that 
their endings contain stretch receptors. These long dendrites provide 
long-range coupling between remote segments of the body, and their 
directionality correlates with the direction of motion. In the circuit
controling backward motion dendrites in A neurons are elongated in the
opposite direction. Ventral side of the worm corresponds to the side
containing cell bodies.

\vspace{0.3cm}

Fig. 3\\
Diagrams showing worm's body and forces acting on a slice of its body 
during undulatory locomotion. All the forces and moments acting on the slice
must balance each other.

\vspace{0.3cm}

Fig. 4\\
Conservation of the normalized wavelength $\lambda/L$ across a population
of wild-type {\it C. elegans}, its mutants, and across related
{\it Caenorhabditis\/} species for forward movement. 
Diamonds represent data points and solid lines are
the least-square fits to the formula $\lambda/L= a_{0} + a_{1}\omega$.
(A) For a population of wild type {\it C. elegans\/} the fit yields:
$a_{0}= 0.629$, $a_{1}= -0.035$ $(N=58)$.
Each data point corresponds to average values of $\lambda/L$ and $\omega$
for one worm.
(B) For different mutants of {\it C. elegans\/} including its wild-type
the fit yields:
$a_{0}= 0.639$, $a_{1}= -0.068$ $(N=17)$. The mutants included:
BE101 {\it sqt-1(sc101)} and BE103 {\it sqt-1(sc103)}, BE109, BC347 
{\it unc-54(s74)}, RW130 {\it unc-54(st130)}, RW132 
{\it unc-54(st132)}, RW134 {\it unc-54(st134)}, RW135 
{\it unc-54(st135)}, and RW5008 {\it unc-54(s95)}, {\it goa-1(n1134)} 
and {\it goa-1(sy192)}, {\it egl-30(tg26)}, 
{\it cat-2(e1112)}, {\it cat-4(e1141)}, CB185 {\it lon-1(e185)}, and 
PS697 {\it lon-1(e185);lon-2(e678)}. The data points are averages over 
a population for each mutation. 
(C) For related {\it Caenorhabditis\/} species the fit yields: 
$a_{0}= 0.659$, $a_{1}= -0.066$ $(N=6)$. 
The species included: N2 {\it C. elegans\/}, 
AF16 {\it C. briggsae\/}, SB339 {\it C. japonica\/}, SB146
{\it C. remanei\/}, CB5161 {\it Caenorhabditis\/} sp., and PS1010
{\it Caenorhabditis\/} sp. The data points are averages over a population for
each species. Note that the normalized wavelength is almost frequency 
independent ($a_{1}$ is close to zero) and very similar in all three figures.

\vspace{0.3cm}

Fig. 5\\
Linear scaling of the velocity of forward propulsion (velocity of the center 
of mass) with the velocity of muscle contraction wave and conservation of the
coefficient $\gamma$. Diamonds represent data points and solid
lines are the least-square fits to Eq. (13) in the main text. The slope
of the regression line corresponds to the average coefficient $\gamma$.
The slope of the dashed lines indicate the maximal allowed slope, i.e., 1.
(A) For a population of wild type {\it C. elegans\/}: 
$\gamma= 0.787$ $(N=58, R^{2}= 0.900)$. Each data point corresponds to 
average values for one worm. (B) For a group of mutants of 
{\it C. elegans\/}, the same as in Fig. 4B, including its wild-type:
 $\gamma= 0.791$ $(N=14, R^{2}= 0.988)$, except for 3 data point (circles) 
representing BE101 {\it sqt-1(sc101)}, BE103 {\it sqt-1(sc103)}, and 
{\it cat-4(e1141)\/} 
mutants that have significantly reduced the ratio $c_{N}/c_{L}$. 
(C) For related {\it Caenorhabditis\/} species, the same as in Fig. 4C, 
the slope is $\gamma= 0.861$ $(N=6, R^{2}= 0.882)$. The data points in (B)
and (C) are averages within a population for each mutation (B) and each
species (C). Note a very similar and almost optimal value of the coefficient
$\gamma$ for all three cases.

\vspace{0.3cm}

Fig. 6\\
Linear scaling of the amplitude of undulations with the wavelength
during different developmental stages. The least-square fit to the
data points yields regression line 
$A_{0}= 0.194\lambda + 0.006$ with $R^{2}=0.98$.

\vspace{0.3cm}

Fig. 7\\
Dependence of the total bending moment $M$ on the position along worm's
body. This dependence has a non-linear character with a travelling wave 
of activity present. Parameters used: $A_{0}= 0.1$ mm, $L= 1.0$ mm,
$\lambda= 0.66$ mm, $\omega= 1.8$ Hz, $\gamma= 0.8$, $c_{N}= 50$ g/(mm s).

\vspace{0.3cm}

Fig. 8\\
Dependence of the amplitude of undulations on the frequency for different
species (A), (B), and {\it C. elegans\/} mutants (C), (D), (E). 
Diamonds are data points
and solid lines are the least-square fits to the formula 
$A_{0}= a/\left[(1+b^{2}\omega^{2})(1+c^{2}\omega^{2})\right]^{1/2}$,
which is equivalent to that represented by Eq. (14).
One data point corresponds to one worm. The parameters of the fits are: 
(A) for AF16 {\it C. briggsae\/} yields $a= 0.121$, $b= 2.903$,
$c= 0.099$ ($N=5$, $R^{2}= 0.934$); (B) for SB339 {\it C. japonica\/} 
$a= 0.201$, $b= 6.207$, $c= 0.001$ ($N=5$, $R^{2}= 0.560$); 
(C) for BE103 {\it sqt-1(sc103)\/} mutants $a= 0.140$, $b= 3.977$, $c=1.466$ 
($N=10$, $R^{2}= 0.830$); 
(D) for BC347 {\it unc-54(s74)\/} mutants $a= 0.155$, $b= 9.478$, $c= 0.040$ 
($N=5$, $R^{2}= 0.871$); (E) for {\it cat-2(e1112)\/} mutants $a= 0.209$, 
$b= 3.673$, $c= 0.004$ ($N=13$, $R^{2}= 0.503$). The fit for BC347 
{\it unc-54(s74)\/} allows us to estimate an effective crossbridge 
dissociation rate, which turns out to be about 2.5 times smaller than in 
the other two {\it C. elegans} mutants: {\it sqt-1(sc103)\/} and 
{\it cat-2(e1112)\/}, which do not alter muscle dynamics.

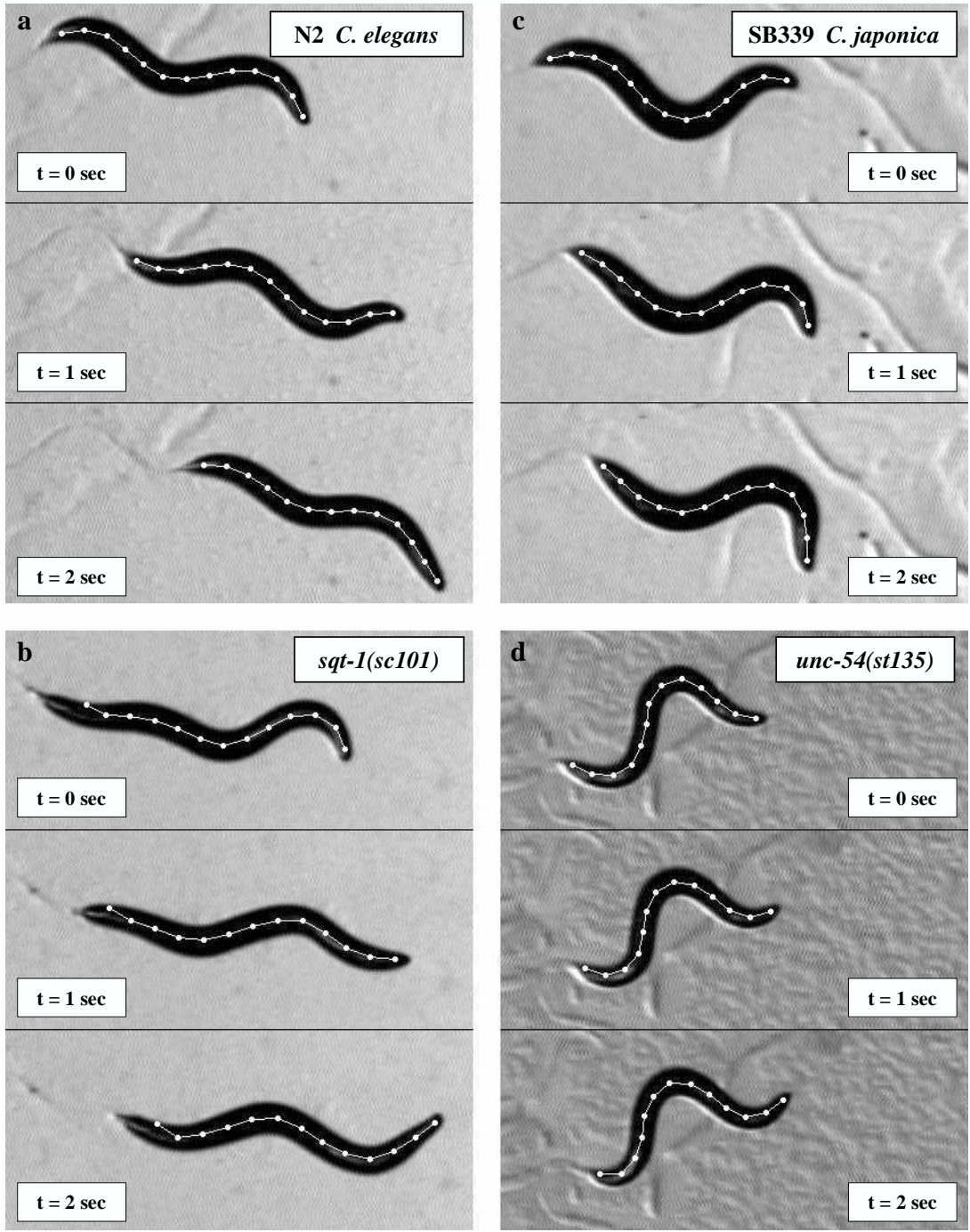
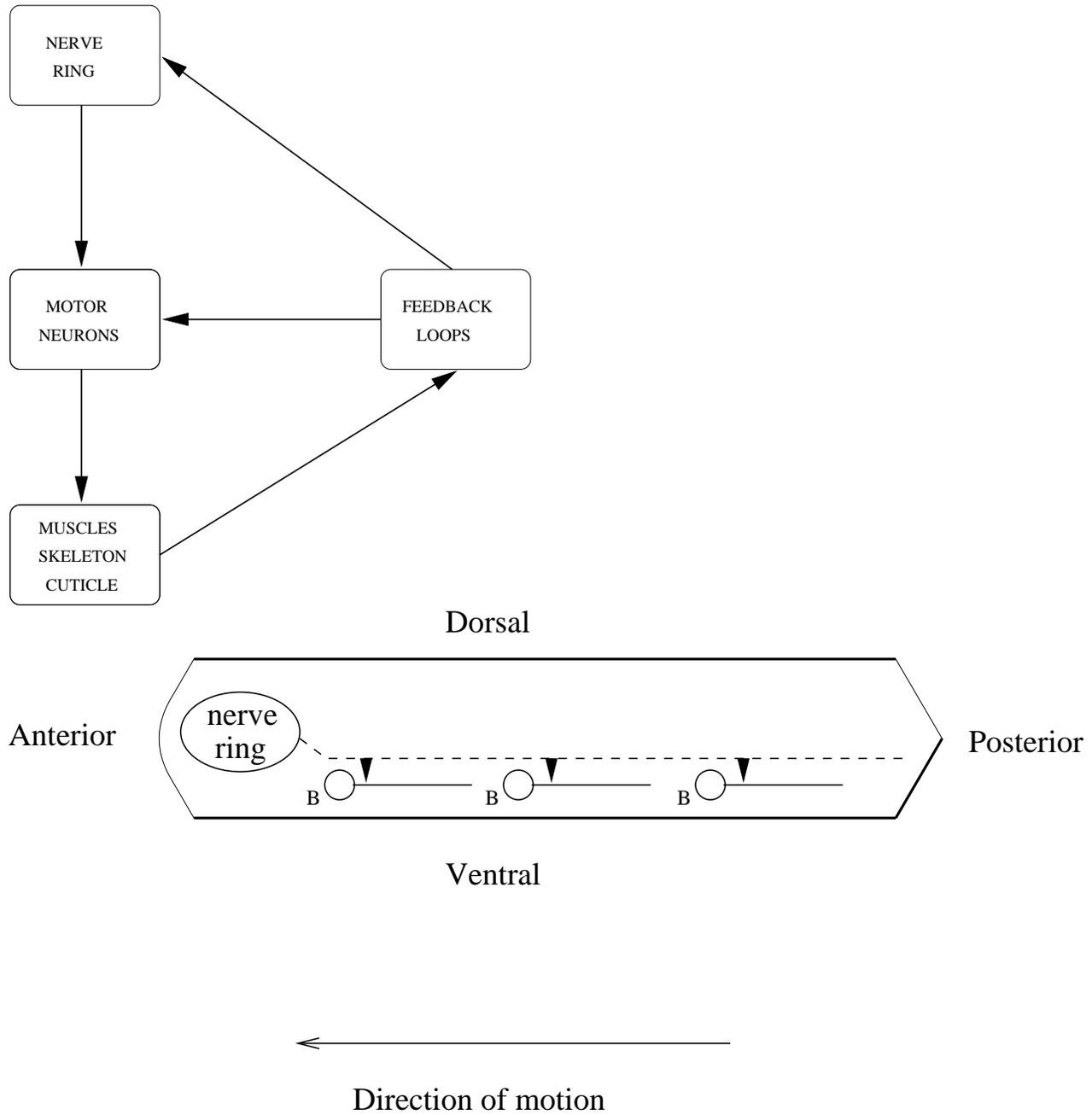
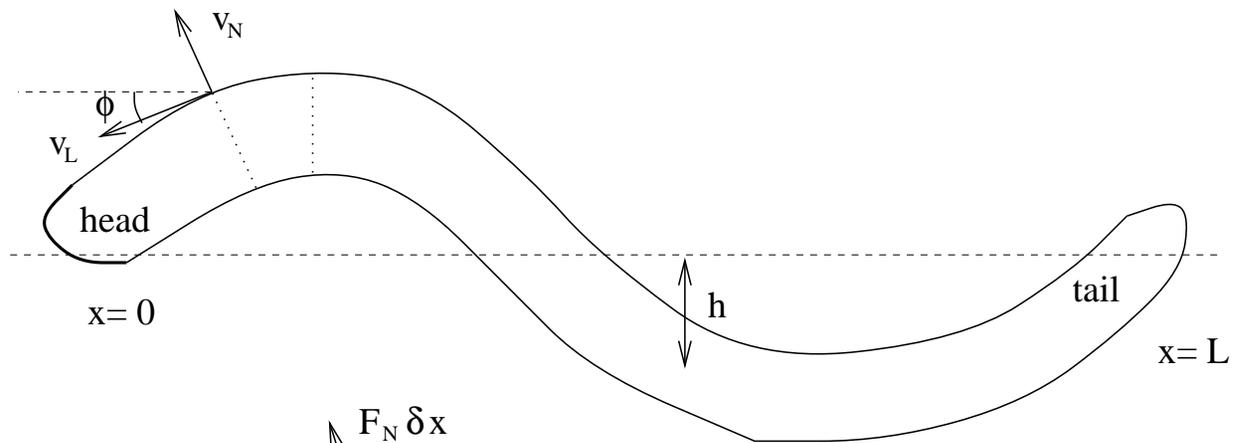
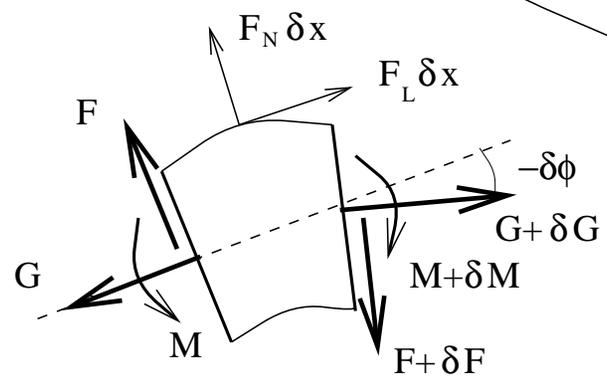
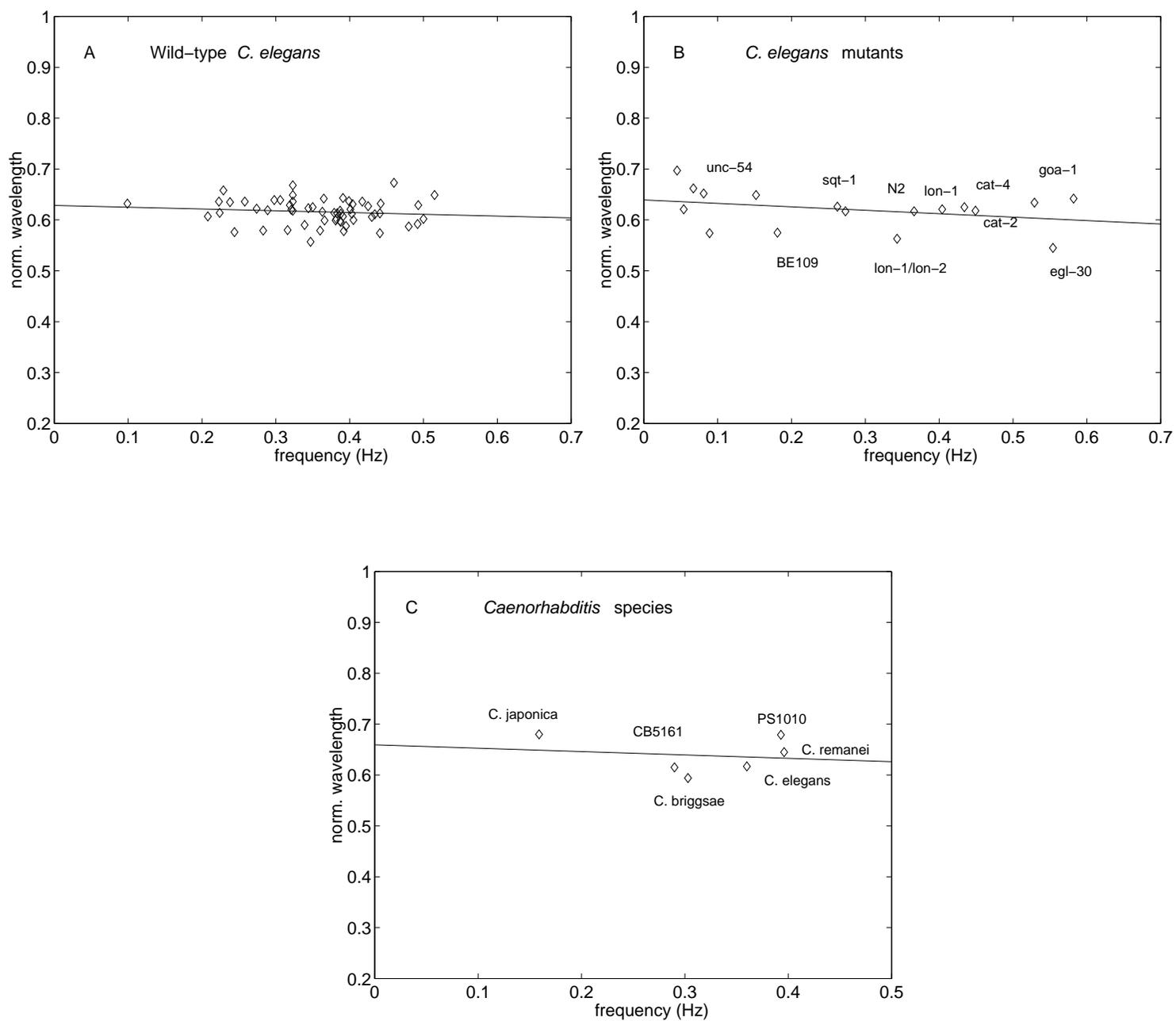
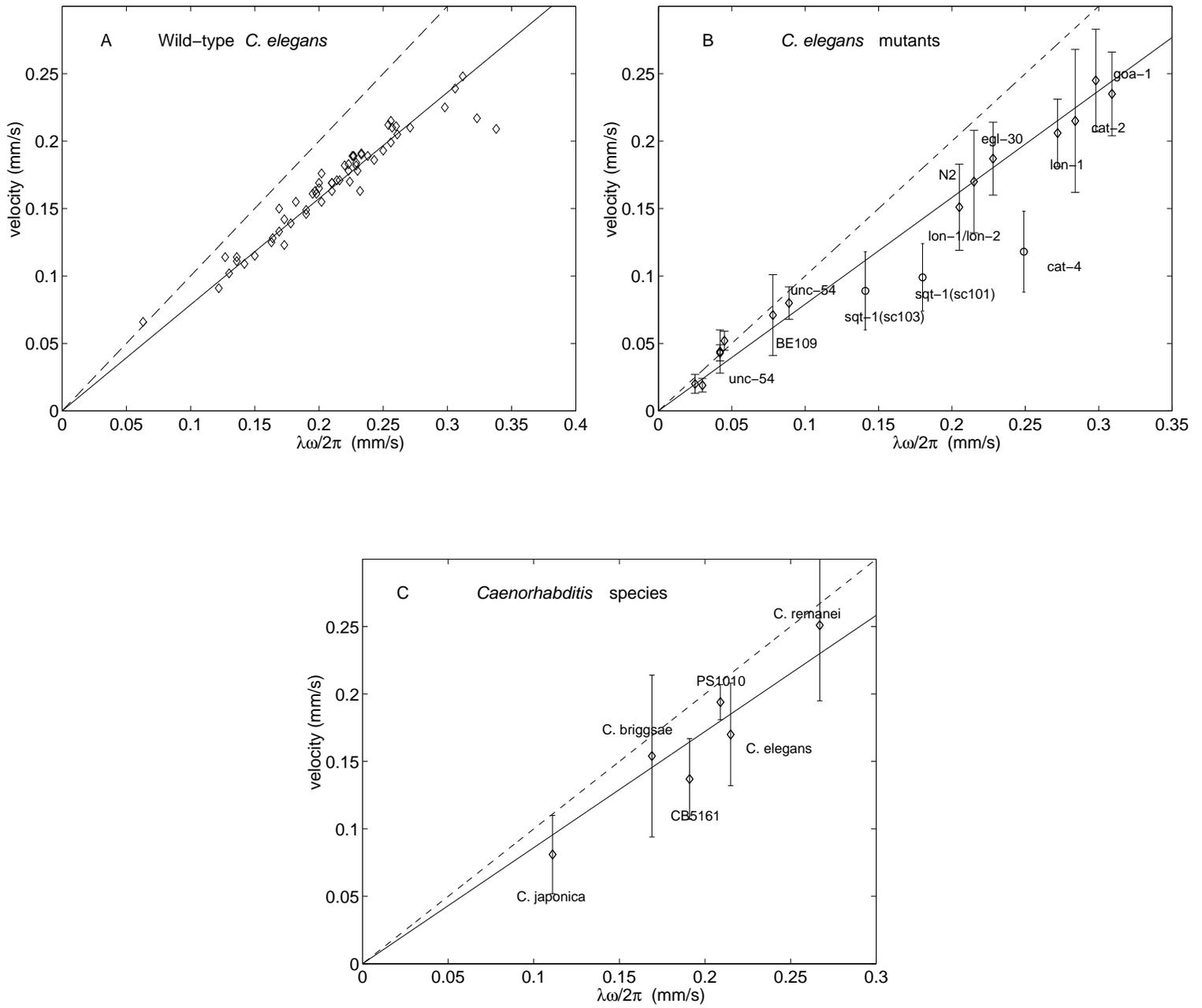
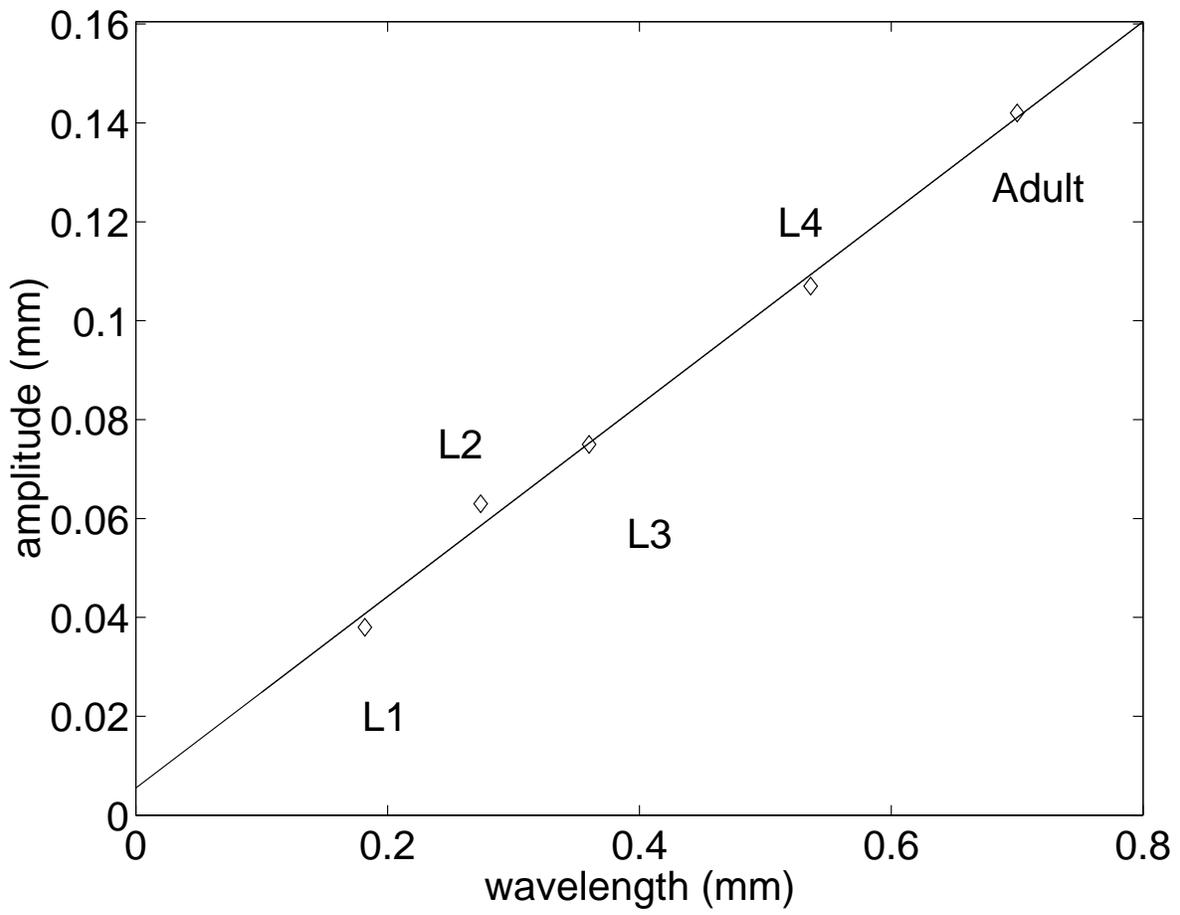
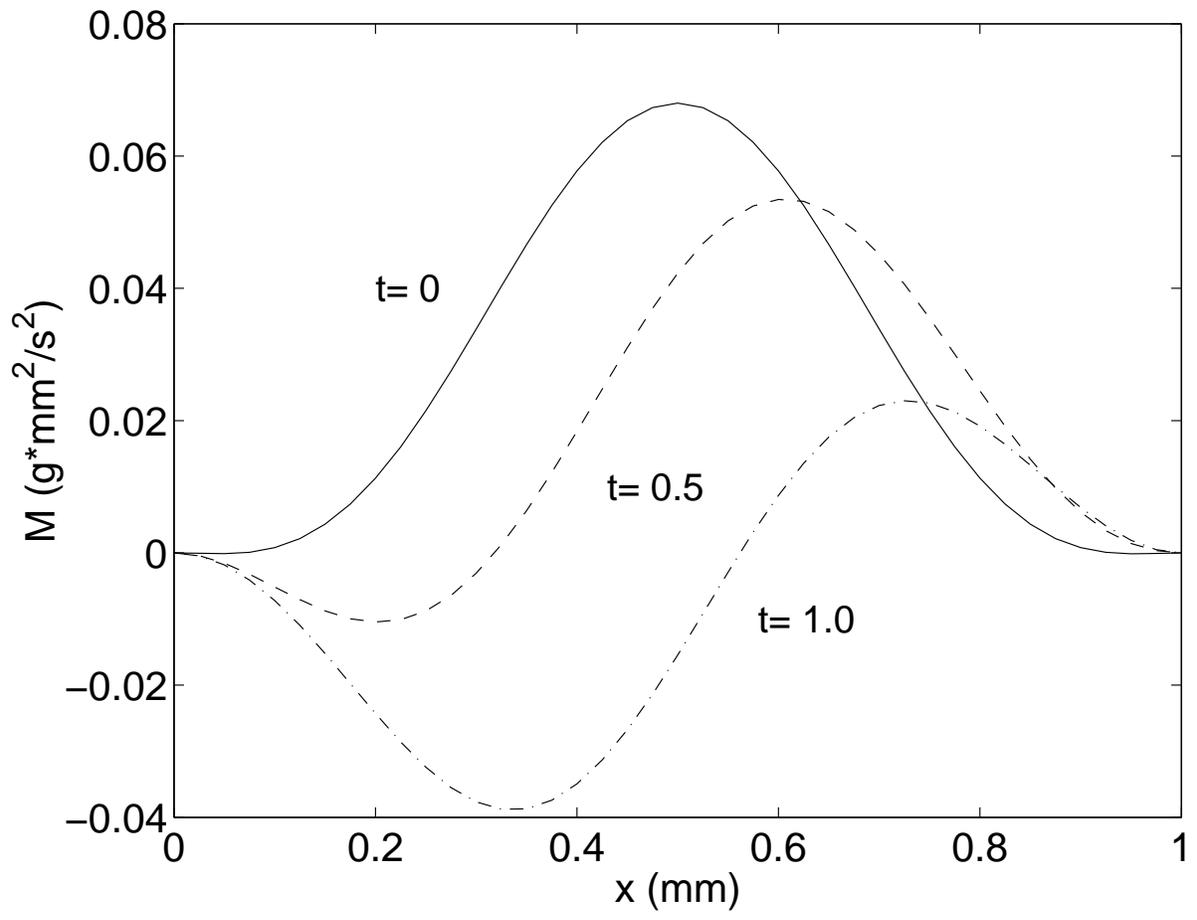
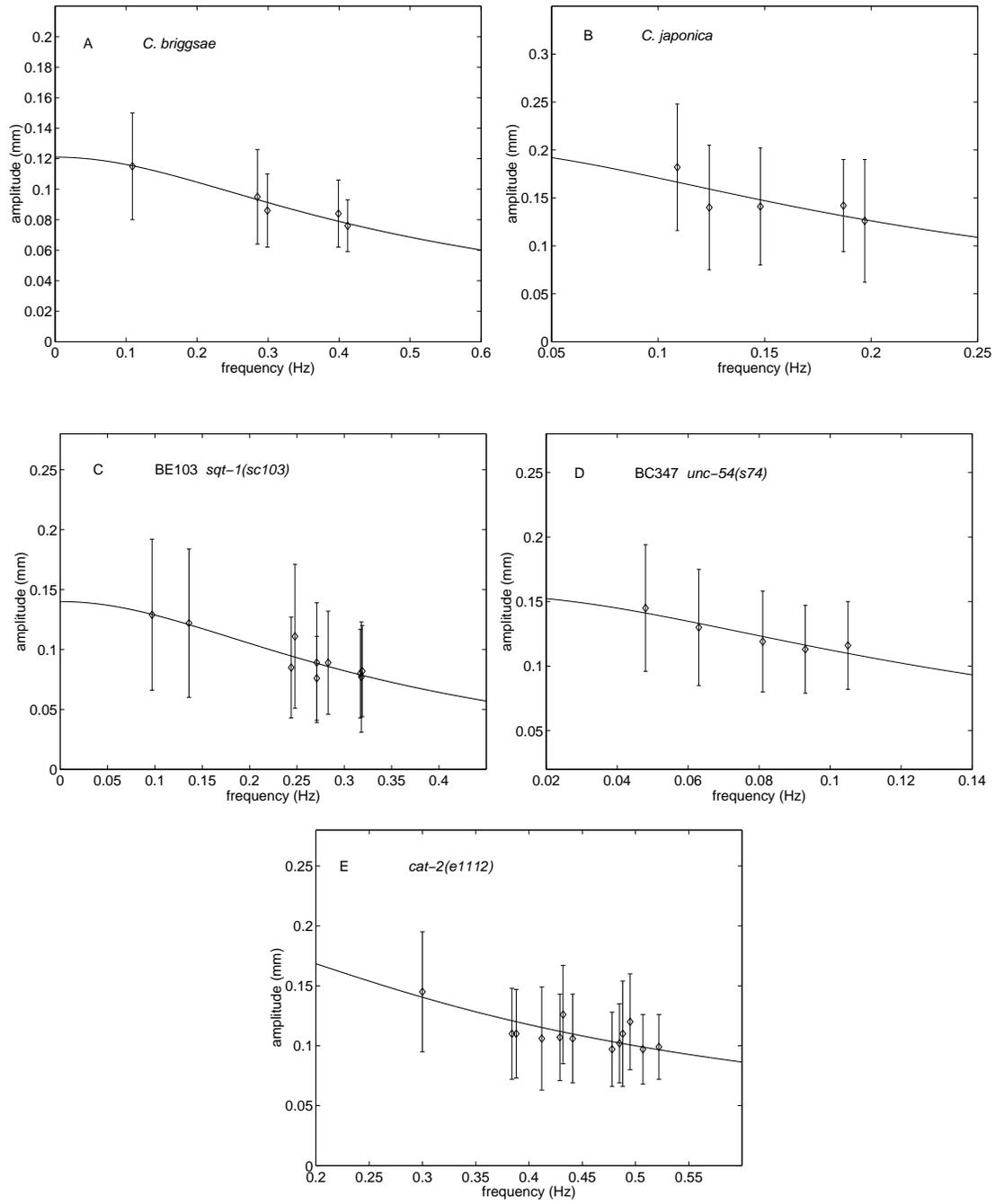
\end{document}